\documentclass[nofootinbib,aps,a4paper,letterpaper,superscriptaddress,
twocolumn,showpacs,times,eqsecnum]{revtex4}
\usepackage{amsmath}
\usepackage{amsfonts}
\usepackage{booktabs}
\usepackage{multirow}
\usepackage{siunitx} 
\usepackage{adjustbox}
\pdfoutput=1
\usepackage{graphicx}
\usepackage{color}
\usepackage{braket}
\usepackage{dcolumn}
\usepackage{bm,url}
\usepackage[linktocpage]{hyperref}
\usepackage{subfigure}
\usepackage{amsfonts}
\usepackage[usenames,dvipsnames,svgnames]{xcolor}  
\usepackage{hyperref}   
\definecolor{oxfordblue}{rgb}{0.0, 0.13, 0.28}
\definecolor{burgundy}{rgb}{0.5, 0.0, 0.13}
\definecolor{darkolivegreen}{rgb}{0.33, 0.42, 0.18}
\definecolor{darkblue}{rgb}{0,0,0.5}
\definecolor{richcarmine}{rgb}{0.84, 0.0, 0.25}
\definecolor{darkblue}{rgb}{0,0,0.5}
\definecolor{bluer}{rgb}{0.00,0.50,0.75}{}
\hypersetup{colorlinks=true, citecolor=red, linkcolor=blue,
 urlcolor = magenta, filecolor=magenta}

\begin{document}
 
 \newcommand\be{\begin{equation}}
  \newcommand\ee{\end{equation}}
 \newcommand\bea{\begin{eqnarray}}
  \newcommand\eea{\end{eqnarray}}
 \newcommand\bseq{\begin{subequations}} 
  \newcommand\eseq{\end{subequations}}
 \newcommand\bcas{\begin{cases}}
  \newcommand\ecas{\end{cases}}
 \newcommand{\p}{\partial}
 \newcommand{\f}{\frac}

 \title{Multi-probe analysis of strong-field effects in $f(Q)$ gravity}
  
 \author{Mohsen Khodadi}
 \email{m.khodadi@du.ac.ir}
 \affiliation{School of Physics, Damghan University, Damghan 3671641167, 
Iran}
 \affiliation{Center for Theoretical Physics, Khazar University, 41 Mehseti 
Str., AZ1096 Baku, Azerbaijan}
 
 \author{Behnam Pourhassan}
 \email{b.pourhassan@du.ac.ir}
 \affiliation{School of Physics, Damghan University, Damghan 3671641167, 
Iran}
 
 \author{Emmanuel N. Saridakis}
 \email{msaridak@noa.gr}
 \affiliation{The National Observatory of Athens, Lofos Nymfon 11852, Greece}
 \affiliation{Departamento de Matem\'{a}ticas, Universidad Cat\'{o}lica del 
  Norte, Avda. Angamos 0610, Casilla 1280, Antofagasta, Chile}
 \affiliation{CAS Key Laboratory for Research in Galaxies and Cosmology, 
School 
  of Astronomy and Space Science,
  University of Science and Technology of China, Hefei 230026, China}

 \begin{abstract}
  Covariant $f(Q)$ gravity is a viable extension of General Relativity, however 
  its strong-field predictions remain largely untested. Using the static,   
spherically symmetric black-hole solutions of the theory, we confront it with   
the most stringent probes available: black-hole shadows, Event Horizon   
Telescope (EHT) measurements, S2-star precession, and strong gravitational 
lensing. We show that the two admissible solution branches behave very 
differently: Case~I produces negligible deviations from Schwarzschild solution, 
whereas   Case~II yields significant, potentially observable corrections to the 
photon   sphere and shadow size. From the EHT shadow diameters of M87* and 
Sgr~A*, we   obtain tight bounds, which are further strengthened by 
strong-lensing coefficients. These results provide the sharpest strong-field 
constraints on   covariant $f(Q)$ gravity to date, and point toward future 
tests using next-generation horizon-scale imaging and precision Galactic-center 
astrometry.
\end{abstract}
 
 \maketitle
 \section{Introduction}

 The discovery of the late-time universe acceleration   has motivated the
 systematic research of modified gravity theories as alternatives to the
 standard $\Lambda$CDM scenario \cite{CANTATA:2021asi, Capozziello:2011et}. 
 Hence, in the literature one can find various extensions and modifications 
of 
 General Relativity (GR). The 
 simplest way to obtain them  is by extending the usual Einstein-Hilbert 
action, 
 resulting in $f(R)$ gravity
 \cite{Starobinsky:1980te, Capozziello:2002rd, DeFelice:2010aj}, 
 $f(G)$ gravity \cite{Nojiri:2005jg, DeFelice:2008wz}, cubic gravity 
 \cite{Asimakis:2022mbe}, Lovelock gravity \cite{Lovelock:1971yv, 
  Deruelle:1989fj}, or scalar-tensor theories such as Horndeski gravity 
 \cite{Horndeski:1974wa, Santos:2023eqp, Santos:2024cvx} and generalized 
Galileon theory \cite{DeFelice:2010nf}.
 Nevertheless, one can 
 start from the equivalent torsional   formulation of gravity,  and extend it 
in 
 similar ways, obtaining $f(T)$ gravity \cite{Cai:2015emx, 
  Linder:2010py, Chen:2010va},    $f(T, T_G)$ gravity 
 \cite{Kofinas:2014owa, Kofinas:2014daa}, and $f(T, B)$ gravity
 \cite{Bahamonde:2015zma, Bahamonde:2016grb},   scalar-torsion theories 
 \cite{Geng:2011aj}, Teleparallel Equivalent of Horndeski theories   
 \cite{Bahamonde:2019shr, 
  Bahamonde:2020cfv, Capozziello:2023foy, Santos:2024zoh, 
DosSantos:2022exb}, etc.

 More recently, Nester~\cite{Nester:1998mp} proposed a third geometric 
 description of gravity, known as  symmetric teleparallel gravity, in 
 which 
 the gravitational interaction is attributed entirely to 
non-metricity, distinct 
 from curvature-based GR and torsion-based teleparallelism.  
 This formulation has attracted considerable interest as an alternative 
foundation  for constructing modified gravity models. Among these, $f(Q)$ 
gravity is applied as one of the usual generalizations 
 \cite{BeltranJimenez:2017tkd,BeltranJimenez:2019tme}, promoting the 
 Lagrangian from its linear   form to a general function $f(Q)$.  
 This extension  improves significantly the theory’s phenomenology and has 
been 
 applied in diverse settings, including cosmological dynamics, astrophysical 
 modeling, and static spherically symmetric spacetimes
 \cite{Anagnostopoulos:2021ydo,Lazkoz:2019sjl,Lu:2019hra,
  Mandal:2020buf, Frusciante:2021sio, 
  Gadbail:2022jco, Khyllep:2021pcu,Barros:2020bgg,Shabani:2023xfn, 
De:2023xua,
  Dimakis:2021gby,Anagnostopoulos:2022gej,Guzman:2024cwa, 
Boiza:2025xpn, Alwan:2025nka, Basilakos:2025olm} (for a  review see 
\cite{Heisenberg:2023lru}).

 A particularly convenient feature of standard $f(Q)$
 gravity is the Coincident Gauge (CG), which allows choosing coordinates in 
which  the affine connection vanishes. However, although the CG simplifies 
calculations significantly, it faces severe limitations in the context of 
spherically symmetric spacetimes: the field equations collapse to linear form, 
preventing  non-trivial $f(Q)$ black hole solutions \cite{Wang:2021zaz}. This 
issue has been
 resolved within the covariant formulation of $f(Q)$ gravity
 \cite{Zhao:2021zab}, where the connection is allowed to be non-vanishing and
 symmetric. The impact of different affine connections on cosmological 
dynamics
 was further studied in \cite{Dimakis:2022rkd,Yang:2024tkw,Ayuso:2025vkc}.

 While $f(Q)$ gravity is strongly motivated by its success in explaining 
 cosmological dynamics, a complete assessment of its viability requires 
 confronting the theory with strong-field observations, where nonlinear 
 gravitational effects dominate. Such environments offer independent and 
 complementary tests to cosmology, since even small modifications of the 
metric 
 can produce measurable deviations near compact objects. Recent efforts have 
 begun to explore this regime through neutron-star structure 
 \cite{Alwan:2024lng}, black-hole solutions \cite{Wang:2024dkn}, and 
 gravitational-wave propagation \cite{Karmakar:2025iba,Karmakar:2025yng}. 
These  studies emphasize that examining the theory under extreme conditions is 
 essential  both for identifying potential departures from GR and for breaking 
degeneracies  that remain unresolved at low curvature.
 
 Among all strong-field systems, black holes provide the most powerful and 
 cleanest probes of high-curvature physics. The horizon-scale images of M87* 
 \cite{EventHorizonTelescope:2019dse,EventHorizonTelescope:2019ggy} and Sgr 
A*  \cite{EventHorizonTelescope:2022wkp} obtained by the Event Horizon 
Telescope 
 (EHT) have opened a new era of precision tests of gravity in the 
immediate 
 vicinity of the photon sphere. These observations have already stimulated a 
 vast 
 body of work exploring horizon-scale phenomenology across different 
 modified-gravity models 
 \cite{Ahmed:2025boj,Ovgun:2025stp,Fathi:2025ikx,AraujoFilho:2025vgb,Xiong:2025wgs,DelPiano:2024nrl,
  Khodadi:2024ubi,Nojiri:2024txy,Liu:2024soc,Hazarika:2024cji,Liu:2024lve,
  Nozari:2024jiz,Sun:2023syd,Nozari:2023flq,Filho:2023ycx,Uniyal:2023ahv,
  Escamilla-Rivera:2022mkc,Khodadi:2022ulo,
  Heidari:2024bkm,Afrin:2022ztr,
  Shaikh:2022ivr,KumarWalia:2022aop,Khodadi:2022pqh,Banerjee:2022jog,
Khodadi:2021gbc,
  Khodadi:2020gns,Khodadi:2020jij,Vagnozzi:2020quf,Vagnozzi:2019apd,
Allahyari:2019jqz,Tsupko:2019pzg}. 
 In the context of covariant $f(Q)$ gravity, static and spherically 
symmetric 
 black-hole spacetimes can be constructed by adopting appropriate 
non-vanishing 
 symmetric connections, as proposed in \cite{DAmbrosio:2021zpm}. However, a 
 systematic confrontation of these solutions with the expanding set of 
 strong-field observables, including EHT shadows, stellar orbital dynamics, 
and 
 strong gravitational lensing, remains absent from the literature.
 
 In this work, we provide the first comprehensive strong-field test of 
power-law 
 $f(Q)$ gravity, which serves as a robust approximation to general viable 
 $f(Q)$ models. We derive and analyze the black-hole solutions associated 
with 
 different symmetric connections, we study their photon-sphere and shadow 
 properties, and we confront them with the EHT measurements of M87* and Sgr~A*. 
We  further extract independent constraints from the periastron precession of 
the  S2 
 star and we examine the predictions for strong gravitational lensing using the 
 Bozza formalism. By combining these complementary probes, we obtain robust 
and  mutually consistent bounds on the deformation parameter $\alpha$, 
allowing us  to  identify which classes of $f(Q)$ black holes remain 
observationally viable in  the strong-gravity regime.

 The remainder of this manuscript is organized as follows.  
Section~\ref{BHsolutions} derives the black hole solutions and their horizon 
structure for the two ansätze of the affine connection. 
Section~\ref{BHShadow} analyzes the photon 
 sphere  and black hole shadow. Section~\ref{EHT} confronts the model with 
EHT  observations of M87* and Sgr A*. Section~\ref{s2} derives constraints 
from the  orbital dynamics of the S2 star. Section~\ref{len} investigates 
strong gravitational lensing and provides further bounds on the model. Finally,
 Section~\ref{con} summarizes our findings and discusses their implications.

 \section{Black hole solutions in $f(Q)$ gravity}
 \label{BHsolutions}

 In this section we summarize the static, spherically symmetric black-hole
 solutions of covariant $f(Q)$ gravity derived in Ref.~\cite{Wang:2024dkn},
 which form the basis for the observational tests carried out in later 
sections.
 We begin with the action,
 \begin{equation}
  S=\frac{1}{16\pi}\int d^4x \sqrt{-g}\, f(Q) + S_{\rm matter},
 \end{equation}
 where $f(Q)$ is a free function of the nonmetricity scalar $Q$, and $S_{\rm 
matter}$ is the matter action. The latter is
 constructed from the nonmetricity tensor
 \[
 Q_{\alpha\mu\nu}
 = \nabla_\alpha g_{\mu\nu}
 = \partial_{\alpha}g_{\mu \nu} 
 - \Gamma^{\lambda}_{\ \alpha\mu}g_{\lambda \nu} 
 - \Gamma^{\lambda}_{\ \alpha\nu}g_{\mu \lambda},
 \]
 and is defined as
 \begin{equation}
  Q = -Q_{\alpha\mu\nu}P^{\alpha\mu\nu},
 \end{equation}
 where $P^{\alpha}_{\ \mu\nu}$ is the nonmetricity conjugate,
 {\small{
 \begin{equation}
  P^{\alpha}_{\ \mu\nu}
  =
  -\frac{1}{4}Q^{\alpha}_{\ \mu\nu}
  +\frac{1}{2}Q_{(\mu\ \nu)}^{\ \ \alpha}
  +\frac{1}{4}\left(Q^{\alpha}-\tilde Q^{\alpha}\right)g_{\mu\nu}
  -\frac{1}{4}\delta^\alpha_{(\mu}Q_{\nu)} ,
 \end{equation}}}
 with traces $Q_\alpha \equiv Q_{\alpha\ \mu}^{\ \mu}$ and $\tilde Q_\alpha 
 \equiv
 Q^{\mu}_{\ \alpha\mu}$.
 
 Varying the action with respect to the metric and the affine connection 
yields
 the field equations,
 \begin{eqnarray}
  &&\!\!\!\!\!\!\!\!\!\!\!\!
  \frac{2}{\sqrt{-g}}\nabla_\alpha
  \left( \sqrt{-g}f_Q P^\alpha_{\ \mu\nu} \right)
  + \frac{1}{2}g_{\mu\nu}f\nonumber\\
  &&
  + f_Q\!\left( P_{\mu\alpha\beta}Q_\nu^{\ \alpha\beta}
  - 2Q_{\alpha\beta\mu}P^{\alpha\beta}_{\ \ \ \nu}\right)
  + 8\pi T_{\mu\nu}=0,
  \label{Eq: metric-eom}
  \\
  && \!\!\!\!\!\!\!\!\!\!\!\! \nabla_\mu\nabla_\nu
  \left( \sqrt{-g}f_Q P^{\mu\nu}_{\ \ \alpha} \right)=0,
 \end{eqnarray}
 where $f_Q\equiv \partial_Q f(Q)$.

 After symmetry reduction of metric and connection imposed by 
 the torsionless and stationary condition (see the details in  
 \cite{DAmbrosio:2021zpm}), we
 obtain the following expressions 
 \begin{eqnarray}
  \label{field-equation-gtt}  
 &&\!\!\!\!\!\!\!\!\!\!\!\!\!\!\!\!
 \partial _{r }g_{tt} 
  =
   - \frac{g_{tt} \left[2 f_{Q} +g_{rr}\left(Q r^{2} f_{Q} -2f_{Q}-f 
    r^{2}\right)\right]}{2 r f_{Q}}\nonumber\\
    &&    \ \ \ + \frac{g_{tt}\left[r^{2} - g_{rr} 
(\Gamma^{r}_{\ \theta 
     \theta})^{2}\right]}{\Gamma^{r}_{\ \theta \theta} r f_{Q}} 
   f_{QQ} \partial _{r}Q
   \, ,  
 \end{eqnarray}
 \begin{eqnarray}
  \label{field-equation-grr}
 &&\!\!\!\!\!\!\!\!\!\!\!\!\!\!
 \partial _{r }g_{rr} 
   = 
   \frac{g_{rr} \left[2 f_{Q} + g_{rr}\left(Q r^{2} f_{Q} -2f_{Q}-f 
    r^{2}\right)\right]}{2 r f_{Q}}
\nonumber\\
    &&   \ \ \ \   
   + \frac{g_{rr}\left[r^{2} + g_{rr} (\Gamma^{r}_{\ \theta 
\theta})^{2} +2 
    r \Gamma^{r}_{\ \theta \theta} \right]}{\Gamma^{r}_{\ \theta 
\theta} r f_{Q}}
   f_{QQ} \partial _{r}Q  ,
 \end{eqnarray}
 \begin{eqnarray}
  \label{nonmetricity-scalar}
&& Q  =   -\frac{1}{r^{2} (\Gamma^{r}_{\ \theta \theta})^{2} g_{tt} 
g_{rr}^{2}}\nonumber \\
&&  \qquad \times
  \Big \{ 
   \partial_{r }g_{tt} \Gamma^{r}_{\ \theta \theta} g_{rr}\left[ 
   g_{rr}(\Gamma^{r}_{\ \theta \theta})^{2} + r (r + 2\Gamma^{r}_{\ 
\theta \theta}) 
   \right]\nonumber\\
&& \qquad \ \ \ \
   + g_{tt} \left[-2 g_{rr}^{2} (\Gamma^{r}_{\ \theta \theta})^{3} 
   \Gamma^{r}_{\ r r} -r^{2} \Gamma^{r}_{\ \theta \theta} \partial_{r} 
   g_{rr}\right] 
\nonumber\\
&&  \qquad \ \ \ \
   + g_{tt} g_{rr} \left[2 r^{2} + 4 r \Gamma^{r}_{\ \theta \theta} + 
2   (\Gamma^{r}_{\ \theta \theta})^{2} \right.
 \nonumber\\
&& \qquad \ \ \ \ \left. + 2 r^{2} \Gamma^{r}_{\ \theta \theta} \Gamma^{r}_{\ 
r r} + 
  (\Gamma^{r}_{\ \theta \theta})^{3} \partial_{r}g_{rr}\right]
  \Big \}  \, ,
 \end{eqnarray}
 where $f_{QQ} \equiv \partial_{QQ}f(Q)$.

 After imposing these symmetries
 \cite{DAmbrosio:2021zpm}, the only nonvanishing independent component of the
 connection becomes
 \begin{equation}\label{eq:Gamma_sign_choice}
  \Gamma^r_{\ \theta\theta}
  = \pm \frac{r}{\sqrt{B(r)}},
 \end{equation}
 which obeys the differential condition
 \begin{equation}
  \partial_r \Gamma^r_{\ \theta\theta}
  = -1 - \Gamma^r_{\ \theta\theta}\Gamma^r_{\ rr}.
  \label{diffeq-Gamma-r-theta-theta}
 \end{equation}
Equation (\ref{diffeq-Gamma-r-theta-theta}) fixes \(\Gamma^{r}_{rr}\) once \(\Gamma^{r}_{\theta\theta}\) is chosen; hence the independent variables in the reduced system are \(A(r), B(r)\), and the non‑metricity scalar \(Q(r)\) which take the form
 \eqref{field-equation-gtt}-\eqref{nonmetricity-scalar}, respectively (see 
 Ref.~\cite{Wang:2024dkn} for the full derivation).

 We consider a static, spherically symmetric metric,
 \begin{equation}
  ds^2 = -A(r)dt^2 + B(r)dr^2
  + r^2(d\theta^2 + \sin^2\theta\, d\varphi^2).
  \label{S}
 \end{equation}
  Concerning the $f(Q)$ form, we will consider the basic model, which is a 
 power-law correction on top of standard general relativity,  namely
 \begin{equation}
  f(Q) = Q + \alpha Q^n - 2\Lambda,
 \end{equation}
 where $\alpha$ controls deviations from GR and $n=2,3$ are the cases of
 interest. Note that the first case, namely   $n=2$, according to Taylor 
 expansion is a good approximation of any realistic $f(Q)$ gravity, i.e. of 
any 
 theory that is a small deviation from general relativity.  
 
 Working perturbatively in $\alpha$, the metric components are expanded
 as
 \begin{equation}
  A(r) = g^{(0)}_{tt} + \alpha g^{(1)}_{tt},
  \qquad
  B(r) = g^{(0)}_{rr} + \alpha g^{(1)}_{rr},
 \end{equation}
 with zeroth-order terms given by the Schwarzschild-de Sitter solution,
 \begin{equation}
  g_{tt}^{(0)} = 1 - \frac{2M}{r} - \frac{\Lambda r^2}{3},
  \qquad
  g_{rr}^{(0)} = 1 + \frac{2M}{r} + \frac{\Lambda r^2}{3}.
 \end{equation}

 Depending on the sign choice in $\Gamma^r_{\ \theta\theta}$, two families of
 solutions arise.\\
 
 \paragraph*{Case I: \(\Gamma^r_{\ \theta\theta} = -r/\sqrt{B(r)}\).}
 \begin{align}
  A(r) &= 1 - \frac{2M}{r} - \frac{\Lambda}{3}r^2
  - \alpha\,\frac{(-1)^n 2^n n M^{2n-1}}{(4n-3) r^{4n-3}},
  \\
  B(r) &= 1 + \frac{2M}{r} + \frac{\Lambda}{3}r^2
  + \alpha\,\frac{(-1)^n 2^n n M^{2n-1}}{r^{4n-3}},
  \\
  Q(r) &= \frac{2M^2}{r^4}
  + \alpha\,\frac{(-1)^n 2^{n+1}nM^2}{r^{4n}}.
 \end{align}
 \\
 \paragraph*{Case II: \(\Gamma^r_{\ \theta\theta} = +r/\sqrt{B(r)}\).}
 \begin{align}
  A(r) &= 1 - \frac{2M}{r} - \frac{\Lambda}{3}r^2
  - \alpha\,\frac{2^{3n-1}}{(2n-3) r^{2n-2}},
  \\
  B(r) &= 1 + \frac{2M}{r} + \frac{\Lambda}{3}r^2
  + \alpha\,\frac{2^{3n-1}(2n^2 -3n +1)}{(2n-3) r^{2n-2}},
  \\
  Q(r) &= \frac{8}{r^2}
  - \alpha\,\frac{2^{3n+1}(n-1)}{r^{2n}}.
 \end{align}
  The two branches exhibit qualitatively different radial corrections: in 
Case~I
 the deviations scale as $r^{3-4n}$, while in Case~II they scale as 
$r^{2-2n}$.
 This distinction will later prove crucial, as only Case~II leads to 
deviations
 large enough to be constrained by current strong-field observations.
 
 Let us proceed by determining the event horizon locations, neglecting the 
 cosmological constant for simplicity. Imposing
 $A(r_h)=0$ we obtain   polynomial equations for $r_h$, which can be 
 solved perturbatively, and thus the horizon corrections can be extracted.\\
 
 \paragraph*{Case I:}
 \begin{align}
  r_h^{(I,n=2)} &\approx 2M\!\left(1+\frac{\alpha}{20M^2}\right),
  \label{I2}
  \\
  r_h^{(I,n=3)} &\approx 2M\!\left(1-\frac{\alpha}{192M^4}\right).
  \label{I3}
 \end{align}\\
 
 \paragraph*{Case II:}
 \begin{align}
  r_h^{(II,n=2)} &\approx 2M\!\left(1+\frac{8\alpha}{M^2}\right),
  \label{II2}
  \\
  r_h^{(II,n=3)} &\approx 2M\!\left(1+\frac{16\alpha}{3M^4}\right).
  \label{II3}
 \end{align} 
 For completeness, in Table~\ref{tab:horizon_results} we list the numerical 
 values of the event horizon location $\alpha$ in the range $[0,0.5]$, in 
Planck 
 mass units. As we can see,  these
 results demonstrate that the Case~II branch exhibits much stronger 
deviations
 from the Schwarzschild radius, a feature that will be essential for deriving
 observational constraints in the following sections.

 \begin{table}[htbp]
  \centering
  \caption{Event horizon radius \(r_h\) for the two solution branches of 
$f(Q)$ gravity model with different values of $\alpha$, and  \(M=1\).}
  \label{tab:horizon_results}
  \begin{tabular}{ccccc}
   \toprule
   \hline
   {\(\alpha\)} & \multicolumn{4}{c}{Event Horizon 
Radius \(r_h\)} \\   \hline
   \cmidrule(lr){2-5}
   & $\!\!\!\!\!\!\!\!\!\!\!$Case I, \(n=2\ \) & Case I, \(n=3\ \) & Case II, 
\(n=2\ \) & Case II, \(n=3\ \) \\    \hline
   \midrule
   0.0 & 2.00000 & 2.00000 & 2.00000 & 2.00000 \\
   0.1 & 2.00494 & 1.99895 & 3.04939 & 2.34040 \\
   0.2 & 2.00975 & 1.99790 & 3.72022 & 2.40510 \\
   0.3 & 2.01444 & 1.99683 & 4.25500 & 2.43243 \\
   0.4 & 2.01900 & 1.99576 & 4.71500 & 2.44762 \\
   0.5 & 2.02344 & 1.99467 & 5.12311 & 2.45714 \\
   \bottomrule    \hline
  \end{tabular}
 \end{table}

Since all strong-field predictions in the following sections rely on the 
perturbative ansatz
\begin{equation}
A(r)=A_{0}(r)+\alpha\, A_{1}(r), \qquad 
B(r)=B_{0}(r)+\alpha\, B_{1}(r),
\label{eq:pert_expansion_check}
\end{equation}
it is essential to verify that the corrections proportional to $\alpha$ remain 
small throughout the region where observational constraints are extracted. In 
the absence of such a check, one cannot guarantee that the perturbative series 
is under control, especially near the photon sphere where the sensitivity to 
deviations from GR is maximal.

A practical way to assess the validity of the expansion is to evaluate the 
dimensionless ratios
\begin{equation}
\epsilon_{A}(r)\equiv 
\left|\frac{\alpha\,A_{1}(r)}{A_{0}(r)}\right|,
\qquad
\epsilon_{B}(r)\equiv 
\left|\frac{\alpha\,B_{1}(r)}{B_{0}(r)}\right|,
\label{eq:eps_def}
\end{equation}
and demand that $\epsilon_{A}(r)$ and $\epsilon_{B}(r)$ remain well below unity 
in the domain of interest. For black-hole shadow and strong-lensing 
observables, the relevant scale is the photon-sphere neighborhood $r\simeq 
r_{c}\approx 3M$. Therefore, we evaluate $\epsilon_{A}(r)$ at $r=3M$ for both 
branches.

\paragraph*{Case I.}
For Case I, the $\alpha$--correction to the lapse function behaves as 
$A_{1}(r)\propto r^{3-4n}$. At $r=3M$, this yields
\begin{equation}
\epsilon^{(I)}_{A}(3M)
\simeq 
 \left|
\alpha\,\frac{k_{n}}{3^{\,4n-3}}\,\frac{1}{M^{2n-2}A_{0}(3M)}
\right|.
\end{equation}
Using $A_{0}(3M)=1-2/3\simeq 1/3$, we obtain the compact estimate
\begin{equation}
\epsilon^{(I)}_{A}(3M)
\approx 
3\left|\frac{\alpha\,k_{n}}{3^{\,4n-3}M^{2n-2}}\right|.
\label{eq:eps_caseI}
\end{equation}
Even for $\alpha/M^{2n-2}$ at the upper end of the observationally allowed 
range, 
$\epsilon^{(I)}_{A}(3M)$ remains $\mathcal{O}(10^{-3})$-$\mathcal{O}(10^{-2})$, 
confirming that the perturbative treatment is self-consistent in this branch.

\paragraph*{Case II.}
In contrast, the Case II solutions produce much larger corrections, since the 
$\alpha$-term scales as $A_{1}(r)\propto r^{2-2n}$, 
which decays more slowly near the photon sphere. Evaluating 
 \eqref{eq:eps_def} at $r=3M$ gives
\begin{equation}
\epsilon^{(II)}_{A}(3M)
\simeq 
3\left|
c_{n}\,\frac{\alpha}{3^{\,2n-2}M^{2n-2}}
\right|.
\label{eq:eps_caseII}
\end{equation}
For $n=2$, this becomes
\begin{equation}
\epsilon^{(II)}_{A}(3M)
\approx 
\mathcal{O}(10)\times\left|\frac{\alpha}{M^{2}}\right|.
\end{equation}
Hence, values as large as $\alpha/M^{2}\sim 10^{-1}$ would already imply 
$\epsilon_{A}^{(II)}(3M)\gtrsim 1$, signalling a breakdown of the perturbative 
expansion. 
This observation explains why Case II deviations grow rapidly with $\alpha$ and 
why the observational constraints derived later naturally restrict 
$\alpha/M^{2n-2}$ to the few-percent level or below.

The above estimates demonstrate that the perturbative series is uniformly well 
behaved for Case I over the entire parameter region considered. For Case II, 
the expansion remains valid only when the dimensionless quantity 
$\alpha/M^{2n-2}$ is sufficiently small, ensuring that all phenomenological 
predictions presented in this work are derived within the consistent domain of 
the perturbative solution.

A notable feature of the covariant formulation of $f(Q)$ gravity is that 
static, spherically symmetric
solutions admit two distinct families determined by the sign choice in the 
single non-vanishing
independent component of the affine connection (\ref{eq:Gamma_sign_choice}).
Although this sign appears at first sight as a mere algebraic freedom, it in 
fact corresponds to two
physically inequivalent realizations of the compatible symmetric connection. 
Because the affine
structure is dynamical in $f(Q)$ gravity, the choice in 
Eq.~\eqref{eq:Gamma_sign_choice} affects
the form of the nonmetricity scalar, the effective gravitational potential, and 
ultimately the
curvature felt by null and timelike geodesics. For clarity, we summarize the 
physical distinctions
between the two branches.

Using Eq.~\eqref{eq:Gamma_sign_choice}, the nonmetricity scalar takes the 
schematic form
\begin{equation}
Q(r) \sim 
\left(\Gamma^{r}{}_{\theta\theta}\right)^{2}\times 
\mathcal{F}[A(r),B(r),A'(r),B'(r)],
\end{equation}
therefore changing the sign of $\Gamma^{r}{}_{\theta\theta}$ does not simply 
flip $Q$, but modifies the 
radial dependence of the subleading contributions entering $\mathcal{F}$. As a 
result, Case I and Case II
yield distinct scaling behaviors of the $\alpha$-dependent corrections. The 
Case~I branch generates
rapidly decaying corrections $\propto r^{3-4n}$, whereas Case~II produces much 
more slowly decaying
terms $\propto r^{2-2n}$. This difference alone already anticipates the 
observational outcomes found
later: Case~I remains extremely close to GR, while Case~II allows sizable 
deviations near the
photon sphere.

The sign in Eq.~\eqref{eq:Gamma_sign_choice} is tied to how the symmetric 
connection accounts for 
the radial change of the two-sphere geometry in spacetime. In GR, the 
Levi-Civita connection fixes
this sign uniquely via metric compatibility. In $f(Q)$ gravity, however, the 
connection is 
independent and compatibility is replaced by a weaker condition involving the 
nonmetricity tensor.
Thus, the two branches correspond to inequivalent ways to encode the extrinsic 
variation of the
angular sector into the affine structure. Since nonmetricity governs how 
lengths and angles vary
under parallel transport, the sign essentially determines the direction in 
which angular intervals
deform as one moves radially.

The two branches lead to different radial corrections in $A(r)$ and $B(r)$:
\begin{align}
\text{Case I:} \qquad & A(r)=A_{0}(r)-\alpha\,\delta A_{I}(r), \\
\text{Case II:} \qquad & A(r)=A_{0}(r)-\alpha\,\delta A_{II}(r),
\end{align}
with $\delta A_{II}(r)$ much larger than $\delta A_{I}(r)$ near $r\sim 
\mathcal{O}(M)$.
Operationally, this means that Case~II modifies the effective gravitational 
potential far more strongly.
This distinction becomes evident in the photon-sphere position, shadow radius, 
and strong-deflection
coefficients computed in later sections.

Obviously, both branches satisfy the full set of field equations derived from 
the variational principle.
Neither violates the symmetry assumptions nor introduces spurious torsion. 
Furthermore, both reduce
smoothly to the Schwarzschild solution as $\alpha\to 0$, indicating that the 
sign choice does not
produce a pathological GR limit. The difference between the branches therefore 
reflects the inherent
flexibility of the affine structure in symmetric teleparallel gravity, where 
multiple connections can
be compatible with the same metric while leading to physically distinct 
phenomenology.

Due to the fact that  Case~I corrections decay rapidly with radius, its 
deviations from GR remain unobservably 
small for astrophysical black holes. By contrast, Case~II exhibits slowly 
decaying $\alpha$-terms
that survive at the photon sphere, thereby allowing shadow distortions and 
strong-lensing effects
within the sensitivity of current and near-future instruments. The branch 
structure of the theory
thus naturally explains why only Case~II yields phenomenology that can be 
meaningfully constrained
with EHT, S2, and strong-field lensing observations.

In summary, the two branches correspond to physically inequivalent realisations 
of the symmetric connection, each leading to qualitatively different 
nonmetricity and strong-field behavior. This distinction underlies all 
observational results in the remainder of the paper and provides a clear, 
geometric interpretation of why only one branch develops sizable departures from 
the Schwarzschild geometry.

 \section{Black hole  shadows} \label{BHShadow}

 In this section we study null geodesics around the black hole solutions
 presented in Sec.~\ref{BHsolutions} and we derive the corresponding 
photon-sphere
 and shadow radii \footnote{Since the matter action is minimally coupled to the metric, photons follow null geodesics of \(g_{\mu\nu}\); there is no direct coupling between the electromagnetic field and the non‑metricity tensor in the present formulation.}. We restrict attention to equatorial motion
 ($\theta=\pi/2$) in the static, spherically symmetric metric~\eqref{S}. For 
null
 geodesics ($ds^2=0$), the timelike and rotational Killing vectors generate 
two
 conserved quantities: the energy $E$ and angular momentum $L$,
 \begin{equation}
  E = A(r)\,\frac{dt}{d\lambda},
  \qquad
  L = r^2\,\frac{d\phi}{d\lambda},
 \end{equation}
 where $\lambda$ is an affine parameter. Hence,
 \begin{equation}
  \frac{dt}{d\lambda} = \frac{E}{A(r)},
  \qquad
  \frac{d\phi}{d\lambda} = \frac{L}{r^2}.
 \end{equation}
 Imposing the null condition $g_{\mu\nu}\dot x^\mu \dot x^\nu = 0$ gives
 \begin{equation}
  -A(r)\left(\frac{dt}{d\lambda}\right)^2
  + B(r)\left(\frac{dr}{d\lambda}\right)^2
  + r^2\left(\frac{d\phi}{d\lambda}\right)^2 = 0,
 \end{equation}
 which, after substituting the above relations, becomes
 \begin{equation}
  B(r)\left(\frac{dr}{d\lambda}\right)^2 + \frac{L^2}{r^2}
  = \frac{E^2}{A(r)}.
  \label{b}
 \end{equation}
 Introducing the impact parameter $b\equiv L/E$ and using
 $dr/d\lambda = (dr/d\phi)\,(d\phi/d\lambda) = (dr/d\phi)\,L/r^2$, we obtain
 \begin{equation}
  \frac{B(r)}{r^4}\left(\frac{dr}{d\phi}\right)^2 + \frac{1}{r^2}
  = \frac{1}{b^2 A(r)}.
  \label{ss}
 \end{equation}
 
 It is convenient to work with the inverse radial coordinate $u\equiv 1/r$.
 Noting that $dr/d\phi = -u^{-2}du/d\phi$, we rewrite Eq.~\eqref{ss} as
 \begin{equation}
  \left(\frac{du}{d\phi}\right)^2
  = \frac{1}{B(u)}\left[\frac{1}{b^2 A(u)} - u^2\right],
  \label{u}
 \end{equation}
 which is the orbit equation for null geodesics. A circular photon orbit
 (the photon sphere) at radius $r_c$ (or $u_c=1/r_c$) is characterized by
 \begin{equation}
  \frac{du}{d\phi}\Big|_{u_c}=0,
  \qquad
  \frac{d^2u}{d\phi^2}\Big|_{u_c}=0.
 \end{equation}
 The first condition implies, from Eq.~\eqref{u},
 \begin{equation}
  \frac{1}{b^2 A(u_c)} - u_c^2 = 0
  \quad\Rightarrow\quad
  b^2 = \frac{1}{u_c^2 A(u_c)} = \frac{r_c^2}{A(r_c)},
 \end{equation}
 so that the critical impact parameter is
 \begin{equation}
  b_c = \frac{r_c}{\sqrt{A(r_c)}}.
  \label{bc}
 \end{equation}
 For an observer at infinity, $b_c$ is precisely the shadow radius $R_s$.
 
 The second condition, $d^2u/d\phi^2=0$, yields the photon-sphere equation. 
 Taking
 the derivative of Eq.~\eqref{u} with respect to $\phi$ gives
 \begin{equation}
  2\frac{du}{d\phi}\frac{d^2u}{d\phi^2}
  =   
\frac{d}{d\phi}\left\{\frac{1}{B(u)}\left[\frac{1}{b^2A(u)}-u^2\right]\right\}.
 \end{equation}
 At $u=u_c$ we have $du/d\phi=0$, hence the right-hand side must vanish, 
which
 leads (after some algebra) to
 \begin{equation}
  \frac{d}{du}\left(\frac{1}{A(u)u^2}\right)\Big|_{u_c} = 0.
 \end{equation}
 In terms of the radial coordinate this is equivalent to
 \begin{equation}
  \frac{d}{dr}\left(\frac{A(r)}{r^2}\right)\Big|_{r_c} = 0
  \quad\Rightarrow\quad
  \frac{A'(r_c)}{A(r_c)} = \frac{2}{r_c},
  \label{rc}
 \end{equation}
 which determines the photon-sphere radius $r_c$. As is evident from
 Eqs.~\eqref{bc} and \eqref{rc}, the lapse function $B(r)$ does not affect 
the
 shadow radius seen by a distant observer.
 
 Thus, to compute the shadow radius for the black hole solutions of
 Sec.~\ref{BHsolutions}, we first solve Eq.~\eqref{rc} perturbatively for 
$r_c$
 and then we substitute the result into Eq.~\eqref{bc}, obtaining the 
 $\alpha$-dependent corrections to $R_s$.
 
 \medskip
 \noindent\textbf{Case I.}
 For the Case~I branch we define
 \begin{equation}
  k_n = \frac{(-1)^n 2^n n}{4n-3},
  \qquad
  p_n = (-1)^n 2^n n,
 \end{equation}
 so that
 \begin{align}
  A(r) &= 1 - \frac{2M}{r} - \alpha k_n \frac{M^{2n-1}}{r^{4n-3}},
  \\
  A'(r) &= \frac{2M}{r^2} + \alpha k_n (4n-3)\frac{M^{2n-1}}{r^{4n-2}}.
 \end{align}
 Introducing the function
 \begin{equation}
  F(r) \equiv rA'(r) - 2A(r),
 \end{equation}
 the photon-sphere condition \eqref{rc} is equivalent to $F(r_c)=0$, with
 \begin{equation}
  F(r) = \frac{6M}{r}
  + \alpha k_n (4n-1)\frac{M^{2n-1}}{r^{4n-3}} - 2.
 \end{equation}
 
 For $n=2$ this reduces to
 \begin{equation}
  6M r_c^4 - 2 r_c^5 + \alpha \frac{28}{5}M^3 = 0.
 \end{equation}
 Setting $r_c = 3M + \epsilon$ and expanding to first order in $\epsilon$ 
(for
 small $\alpha$) yields
 \begin{equation}
  \epsilon = \frac{14\alpha}{405M} \approx 0.034568\,\frac{\alpha}{M},
 \end{equation}
and thus
 \begin{equation}
  r_c^{(I,n=2)} \approx 3M + 0.03457\,\frac{\alpha}{M}.
 \end{equation}
 Substituting this into Eq.~\eqref{bc}, the corresponding shadow radius 
becomes
 \begin{equation}
  R_s^{(I,n=2)} \approx 3\sqrt{3}\,M
  \left(1 + 0.00494\,\frac{\alpha}{M^2}\right).
 \end{equation}
 
 Repeating the same procedure for $n=3$ we obtain
 \begin{equation}
  r_c^{(I,n=3)} \approx 3M - 0.002236\,\frac{\alpha}{M^3},
 \end{equation}
 and
 \begin{equation}
  R_s^{(I,n=3)} \approx 3\sqrt{3}\,M
  \left(1 - 0.000203\,\frac{\alpha}{M^4}\right).
 \end{equation}
 
 \medskip
 \noindent\textbf{Case II.}
 For the Case~II branch, the photon-sphere equation simplifies considerably. 
  For $n=2$ one finds
 \begin{equation}
  F(r) = r^2 - 3Mr - 64\alpha,
 \end{equation}
 and imposing $F(r_c)=0$ yields, to first order in $\alpha$,
 \begin{equation}
  r_c^{(II,n=2)} \approx 3M + 21.333\,\frac{\alpha}{M},
 \end{equation}
 with shadow radius
 \begin{equation}
  R_s^{(II,n=2)} \approx 5.19615\,M
  \left(1 + 5.333\,\frac{\alpha}{M^2}\right).
 \end{equation}
 
 For $n=3$ we obtain
 \begin{equation}
  r_c^{(II,n=3)} \approx 3M - 1.8963\,\frac{\alpha}{M^3},
 \end{equation}
 and
 \begin{equation}
  R_s^{(II,n=3)} \approx 3\sqrt{3}\,M
  \left(1 + 1.580\,\frac{\alpha}{M^4}\right).
 \end{equation}
 
 These expressions show that, while the Case~I corrections to the shadow 
radius
 are relatively small, the Case~II branch can lead to significantly larger
 deviations from the GR value, especially for $n=2$. This will make Case~II
 the primary target of the observational constraints derived from EHT, 
S2-star
 dynamics, and strong lensing in the following sections.

 \begin{table}[htbp]
  \centering
  \caption{Photon sphere radius \(r_c\) and shadow radius \(R_s\) for 
  the two solution
branches   of covariant \(f(Q)\) gravity. The dimensions of the  parameter 
\(\alpha\)are: for \(n=2\), \([\alpha] = M^2\); for \(n=3\), 
\([\alpha] = M^4\), where \(M\) is the BH mass.}
  \label{tab:shadow_results}
  \begin{tabular}{cccc}
     \hline
   \toprule 
  Case & \(n\) & \multicolumn{1}{c}{\(\ \ \ \ \ \  r_c\) (photon sphere)} & 
\multicolumn{1}{c}{\(\ \ \ \ R_s\) (shadow radius)} \\     \hline
   \midrule
   I    & 2 & \(3M + 0.03457 \dfrac{\alpha}{M}\) 
   & \(3\sqrt{3} M \left(1 + 0.00494 \dfrac{\alpha}{M^2}\right)\)      
\vspace{0.2cm}    \\
   \vspace{0.2cm}   
   I    & 3 & \(3M - 0.002236 \dfrac{\alpha}{M^3}\) 
   & \(3\sqrt{3} M \left(1 - 0.000203 \dfrac{\alpha}{M^4}\right)\) \\
      \vspace{0.2cm}   
   II   & 2 & \(3M + 21.333 \dfrac{\alpha}{M}\) 
   & \(3\sqrt{3} M \left(1 + 5.333 \dfrac{\alpha}{M^2}\right)\) \\
      \vspace{0.1cm}   
   II   & 3 & \(3M - 1.8963 \dfrac{\alpha}{M^3}\) 
   & \(3\sqrt{3} M \left(1 + 1.580 \dfrac{\alpha}{M^4}\right)\) \\
   \bottomrule
      \hline
  \end{tabular}
 \end{table}

 In
 Table~\ref{tab:shadow_results} we present an overview of the shadow 
and photon-sphere results. Since the deviation parameter $\alpha$ 
is
 assumed to be small, we compute the photon-sphere radius $r_c$ and the 
shadow
 radius $R_s$ numerically for $\alpha \in [0,0.5]$, with the results 
displayed in
 Table~\ref{tab:shadow_results_single}. As the Tables show, the Case~I branch
 remains practically indistinguishable from the GR prediction, with 
corrections
 well below current observational sensitivity. In contrast, the Case~II 
branch
 exhibits clear and monotonic deviations from GR: increasing $\alpha$ leads 
to a
 larger shadow size, making these solutions observationally testable with
 horizon-scale imaging.

  \begin{table}[htbp]
  \centering
  \caption{Photon sphere radius \(r_c\) and shadow radius \(R_s\)   for 
  the two solution
branches   of covariant \(f(Q)\) gravity,  with different values of 
$\alpha$, and \(M=1\).}
  \label{tab:shadow_results_single}
  \adjustbox{max width=\textwidth}{%
   \begin{tabular}{ccccccccc}
    \toprule
        \hline
   \(\alpha\)  & \multicolumn{2}{c}{Case I, 
\(n=2\)} & \multicolumn{2}{c}{Case I, \(n=3\)} & \multicolumn{2}{c}{Case II, 
\(n=2\)} & \multicolumn{2}{c}{Case II, \(n=3\)} \\
    \cmidrule(lr){2-3} \cmidrule(lr){4-5} \cmidrule(lr){6-7} 
\cmidrule(lr){8-9}     \hline
    & \(r_c\) & \(R_s\) & \(r_c\) & \(R_s\) & \(r_c\) & \(R_s\) & 
\(r_c\) & \(R_s\) \\     \hline
    \midrule
    0.0 & 3.0000 & 5.1962 & 3.0000 & 5.1962 & 3.0000 & 5.1962 
& 3.0000 & 5.1962 \\
    0.1 & 3.0035 & 5.1987 & 2.9998 & 5.1960 & 5.1333 & 6.9675 
& 2.8104 & 6.0168 \\
    0.2 & 3.0069 & 5.2013 & 2.9996 & 5.1959 & 7.2666 & 8.7388 
& 2.6207 & 6.8374 \\
    0.3 & 3.0104 & 5.2039 & 2.9993 & 5.1958 & 9.3999 & 10.5101 
& 2.4311 & 7.6580 \\
    0.4 & 3.0138 & 5.2064 & 2.9991 & 5.1957 & 11.5332 & 
12.2814 & 2.2415 & 8.4787 \\
    0.5 & 3.0173 & 5.2090 & 2.9989 & 5.1956 & 13.6665 & 
14.0527 & 2.0519 & 9.2993 \\
    \bottomrule
    \hline
   \end{tabular}%
  }
 \end{table}

\begin{figure}[ht!]
  \centering
  \includegraphics[width=0.6 \columnwidth]{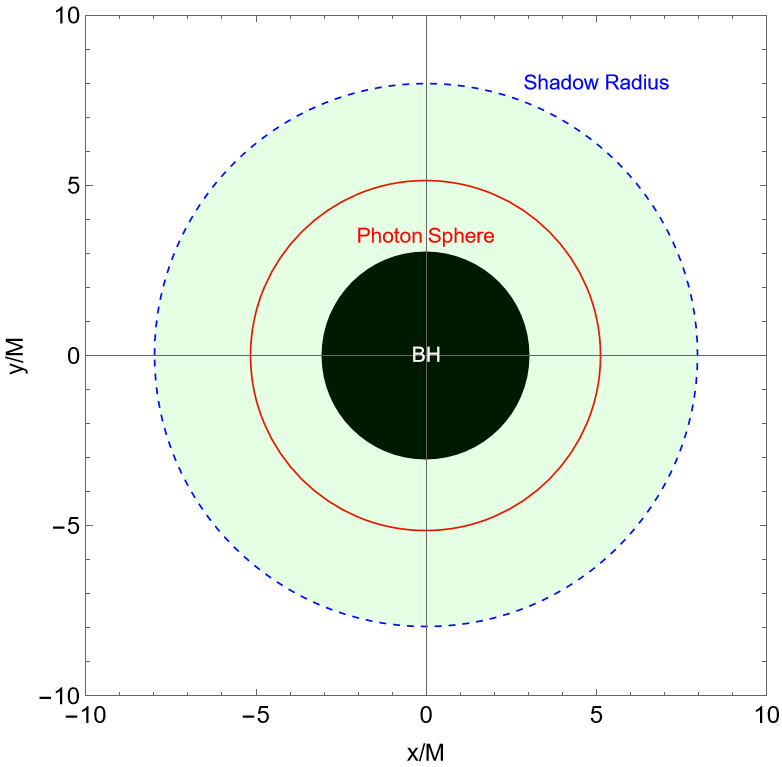}\\
  \includegraphics[width=0.6 \columnwidth]{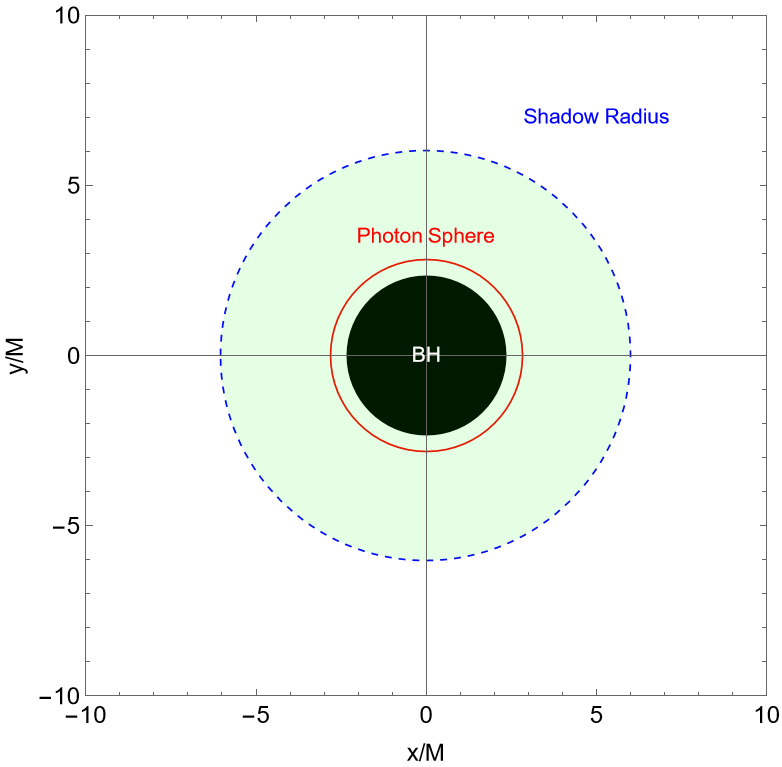}
  \caption{{\it{The black hole shadow of Case II solution in the equatorial 
plane with $n=2$ (upper panel), and $n=3$ (lower panel) for fixed value of 
$\alpha=0.1$ in Planck mass units.} }}
  \label{f1}
 \end{figure}

\begin{figure}[ht!]
  \centering
  \includegraphics[width=0.8 \columnwidth]{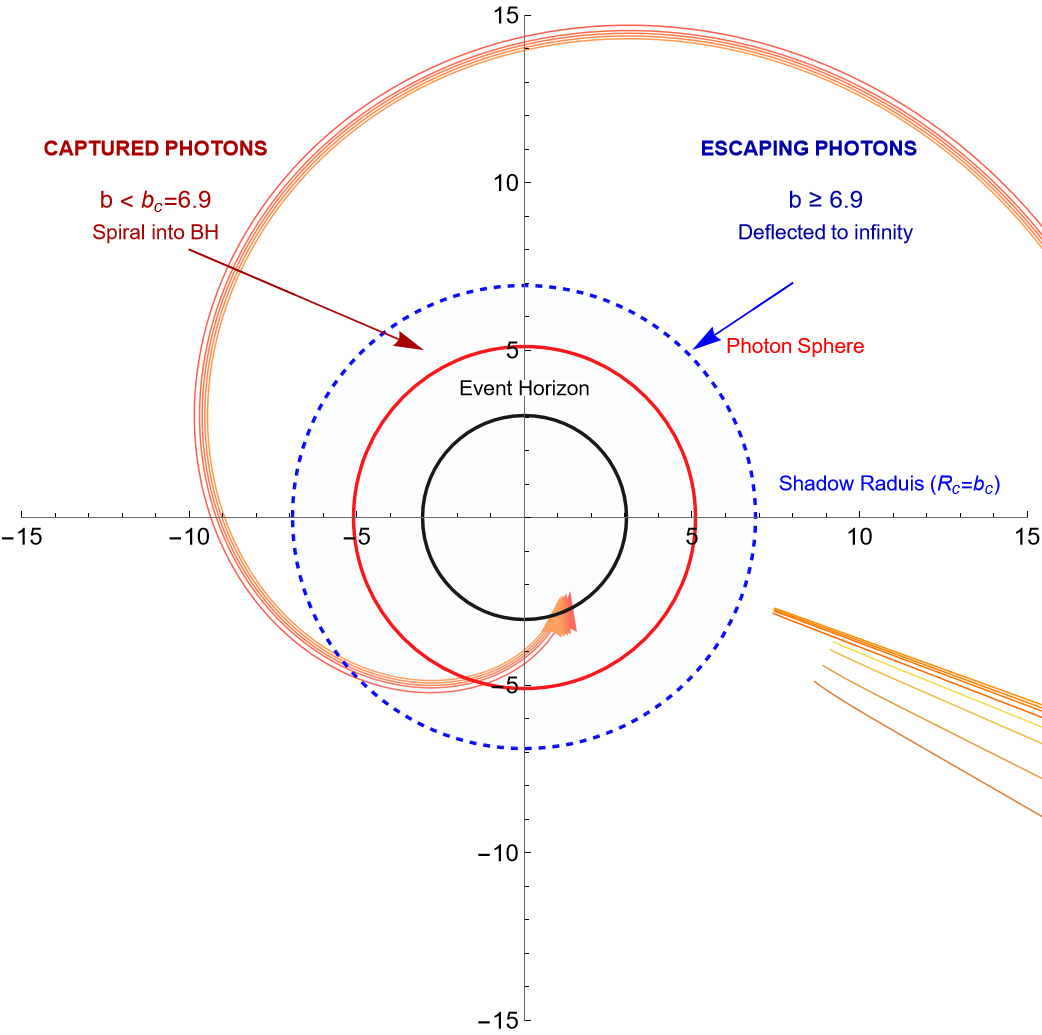}~~
  \caption{{\it{ A representative example of null geodesics in the equatorial 
plane 
of the Case 
II black-hole solution with $n=2$ and $\alpha=0.1$ in Planck mass units.} }}
  \label{f2}
 \end{figure}

 Fig.~\ref{f1} illustrates the corresponding shadow profiles for the 
Case~II
 family with $n=2$ and $n=3$, where we can observe  the 
significant dependence of the
 shadow radius on the deformation parameter. To visualize photon 
trajectories,
 Fig.~\ref{f2} shows sample equatorial null geodesics for the representative 
case
 $n=2$ and $\alpha=0.1$. Photons with impact parameter $b<b_c \simeq 6.9$ are
 captured by the black hole, while those with $b>b_c$ escape to infinity, 
thus
 forming the boundary of the observable shadow.

Before turning to observational tests, it is helpful to provide an intuitive 
explanation of how
nonmetricity alters the propagation of light in covariant $f(Q)$ gravity. 
Although the technical
derivation in the previous parts follows directly from the modified field 
equations, the
underlying physical picture can be summarized in a simple way.

In general relativity, the Levi-Civita connection preserves both lengths and 
angles during parallel
transport. In symmetric teleparallel gravity, the connection is torsionless and 
curvature-free, but
it does not preserve the metric. The nonmetricity tensor 
$
Q_{\alpha\mu\nu}=\nabla_{\alpha}g_{\mu\nu}$
measures precisely how lengths and angles change when transported across 
spacetime. This change
is usually negligible far from compact objects, however it becomes amplified in 
strong fields, where the
geometry varies rapidly.

In $f(Q)$ gravity the geodesics of photons remain the extremals of the metric, 
not of the affine connection. However, the metric is no longer governed solely 
by Einstein's equations, and its radial
profile is affected by the nonmetricity scalar $Q$. The result is that null 
geodesics, although defined in the usual way by $ds^{2}=0$, respond to a 
spacetime geometry whose shape has
been distorted by the nonmetricity-driven corrections. That is why the effect 
is conceptually distinct from modified-optics or birefringent theories: the 
photons behave normally, but the stage
on which they move has been subtly reshaped.

The practical consequence of this geometric reshaping can be seen more easily  
 near the photon sphere.
A small deviation in the metric component $A(r)$ near $r\simeq 3M$ can shift 
the condition (\ref{rc}) which determines the radius of the unstable circular 
orbit of photons. Since the nonmetricity
corrections in the Case~II branch decay slowly with radius, they remain 
relevant at the photon
sphere and can either strengthen or weaken the effective gravitational 
potential felt by null rays.
In turn, this shifts both the location of the light ring and the apparent size 
of the shadow.

Light rays that just graze the photon sphere are exponentially sensitive to the 
shape of the
effective potential. A small nonmetricity-induced deformation of $A(r)$ 
therefore produces
compensated changes in the critical impact parameter $b_{c}$, the size and 
displacement of the shadow boundary, the strong-deflection coefficients $\bar{a}$ 
and $\bar{b}$, and the separation of relativistic images in strong lensing.

The slow decrease of the Case~II corrections implies that  these effects 
survive into the regime probed by
the Event Horizon Telescope, while the more rapidly decaying Case~I corrections 
become effectively
invisible at astrophysical scales. In summary, nonmetricity modifies the 
background geometry rather than the photon propagation rule
itself. This subtle but important distinction explains why $f(Q)$ gravity 
predicts distinctive
signatures in light deflection and black-hole shadow observables while 
remaining compatible with
the geometric-optics limit of photon transport.

 \section{Constraints from Event Horizon Telescope Observations}
 \label{EHT}
 
 The strong-field signatures of black hole spacetimes offer a powerful probe 
of  modified gravity theories. Among these, the Event Horizon Telescope (EHT) 
 observations of the supermassive black holes M87* 
 \cite{EventHorizonTelescope:2019dse,EventHorizonTelescope:2019ggy} and 
Sgr~A*  \cite{EventHorizonTelescope:2022wkp} provide the most precise 
horizon-scale  measurements to date. The EHT effectively resolves the lensed 
photon orbit  surrounding the event horizon, allowing the extraction of an 
angular diameter  associated with the shadow, a quantity that depends solely on 
the near-horizon  geometry, and therefore offers a direct test of the metric. In 
this section we  use the EHT shadow measurements to place observational 
constraints on the  deformation parameter $\alpha$ appearing in   Case~II 
black-hole solutions of    $f(Q)$ gravity.

 The EHT is a global very-long-baseline interferometer (VLBI) operating at 
 230~GHz, achieving an angular resolution of order $\sim 20\,\mu$as. At this 
 wavelength, the emission surrounding the black hole becomes optically thin, 
and 
 the interferometric baselines are long enough to reconstruct the brightness 
 distribution near the photon ring. The observable extracted by the EHT 
 collaboration is an intensity-weighted ring diameter, which closely tracks 
the  theoretical shadow radius $R_s$ for any metric whose near-horizon 
structure is  approximately circular and stationary.
 
 While numerical radiative-transfer models contribute to the detailed 
 interpretation of the images, several studies 
 \cite{Bambi:2019tjh,Vagnozzi:2022moj} have shown that the inferred shadow 
 diameter can be treated with minimal model dependence. This allows for a 
direct 
 comparison between the theoretical shadow radius of a given black hole 
solution 
 and the observationally inferred values. In particular, one can express the 
 measured angular shadow diameter $\theta_{\rm sh}$ as
 \begin{equation}
  \theta_{\rm sh} = \frac{2 R_s}{D},
 \end{equation}
 where $R_s$ is the theoretical shadow radius in units of mass $M$, and $D$ 
is  the distance to the source. Given independent estimates of $D$ and $M$, the 
EHT data thereby place constraints on any theory that modifies $R_s$ relative 
to its  GR value $R_s^{\rm GR}=3\sqrt{3}M$.
 
  As demonstrated in Fig.~\ref{f1} and in 
Table~\ref{tab:shadow_results_single}, 
 the Case~II solutions exhibit a substantially stronger dependence on the 
 deformation parameter~$\alpha$ than the Case~I solutions. This is a direct 
 consequence of the different radial scaling of the $\alpha$-corrections in 
the 
 two families of solutions: Case~II introduces corrections of order $\sim 
 r^{2-2n}$, which decay more slowly than the $\sim r^{3-4n}$ corrections of 
 Case~I, and therefore remain non-negligible near the photon sphere ($r_c 
\simeq 
 3M$). This enhanced sensitivity makes Case~II a natural target for 
observational 
 constraints.
 
   \begin{figure}[ht!]
  \centering
  \includegraphics[width=0.75 \columnwidth]{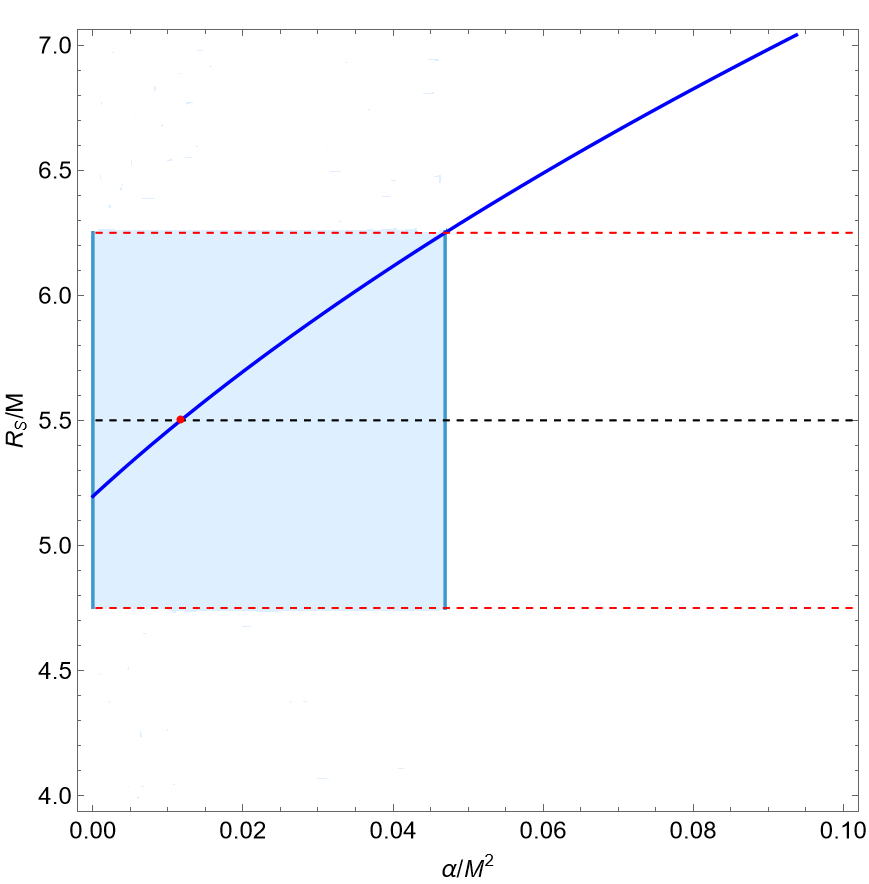}\\
  \includegraphics[width=0.75 \columnwidth]{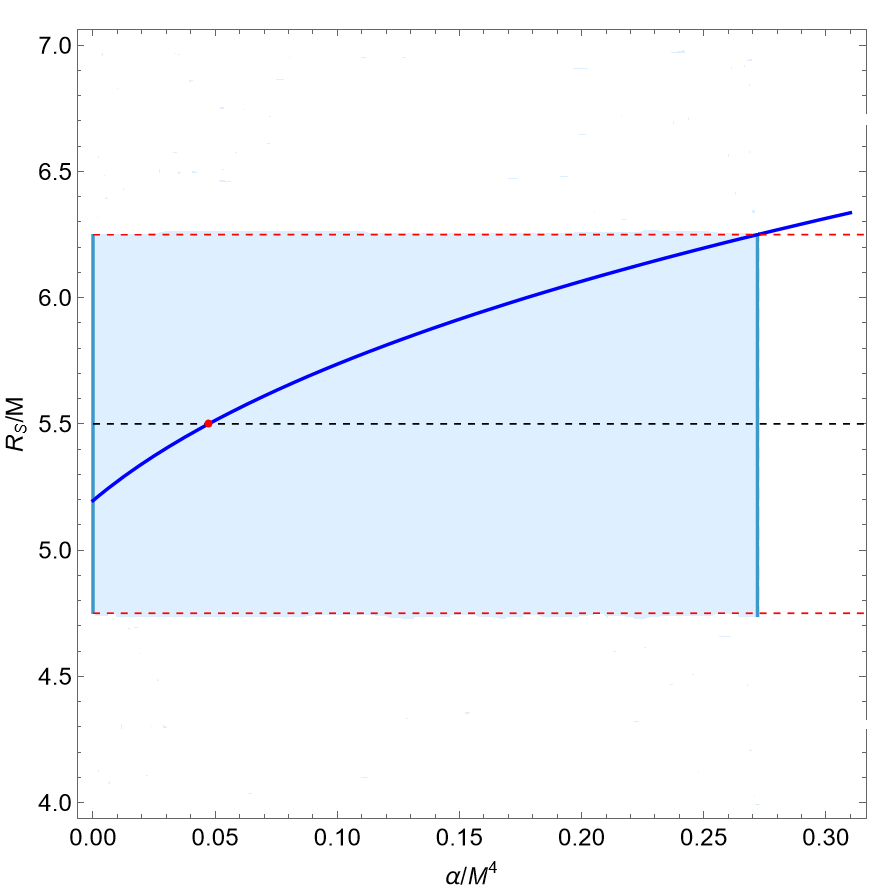}
  \caption{{\it{The allowed range of $\alpha/M^2$ (upper panel), and 
$\alpha/M^4$ 
(lower panel) for Case II black-hole solution with $n=2, 3$ in light of EHT 
diameter measurement of M87*. The black-dashed line represents the central 
value of $R_s/M$ in EHT data, while the red-dashed lines mark the lower and 
upper permissible bounds.} }}
  \label{f3}
 \end{figure}

 For definiteness, we summarise the leading corrections:
 \begin{align}
  R_s^{\rm (II)}(n=2) &\approx 3\sqrt{3} M \left(1 + 
  5.333\,\frac{\alpha}{M^2}\right), \\
  R_s^{\rm (II)}(n=3) &\approx 3\sqrt{3} M \left(1 + 
  1.580\,\frac{\alpha}{M^4}\right).
 \end{align}
 These expressions show that even for $\alpha \sim 10^{-2} M^2$-$10^{-1} 
M^2$, 
 the shift in $R_s$ can reach several percent, which is within the 
sensitivity of 
 current EHT measurements. This motivates confronting the model with 
 observations.
 
 \subsection{Constraints from the M87* shadow}
 
 For M87*, the EHT reports an angular ring diameter of $\theta_{\rm ring} 
\approx 
 42 \pm 3\,\mu{\rm as}$, corresponding to a dimensionless shadow size of
 \begin{equation}
  \frac{2 R_s}{M} \approx 11.0 \pm 1.5,
 \end{equation}
 after combining the angular size with the distance \cite{Bambi:2019tjh}.

 Using the modified shadow radii of the Case~II black hole, we compare the 
 predicted shadow diameter with the allowed observational band. The results, 
 illustrated in Fig.~\ref{f3}, show that the model remains compatible with 
the 
 EHT measurement provided the deformation parameter satisfies
 \begin{align}
  0 < \frac{\alpha}{M^2} &< 0.047 \qquad (n=2), \\
  0 < \frac{\alpha}{M^4} &< 0.27 \qquad (n=3).
 \end{align}
 Larger values of $\alpha$ would increase the shadow radius beyond the upper 
 observational bound, thereby being ruled out. The constraints are therefore 
 primarily upper bounds, with M87* providing the most stringent restrictions 
for 
 $n=2$.
 
 \subsection{Constraints from the Sgr~A* shadow}
 
\begin{figure}[ht!]
  \centering
  \includegraphics[width=0.75 \columnwidth]{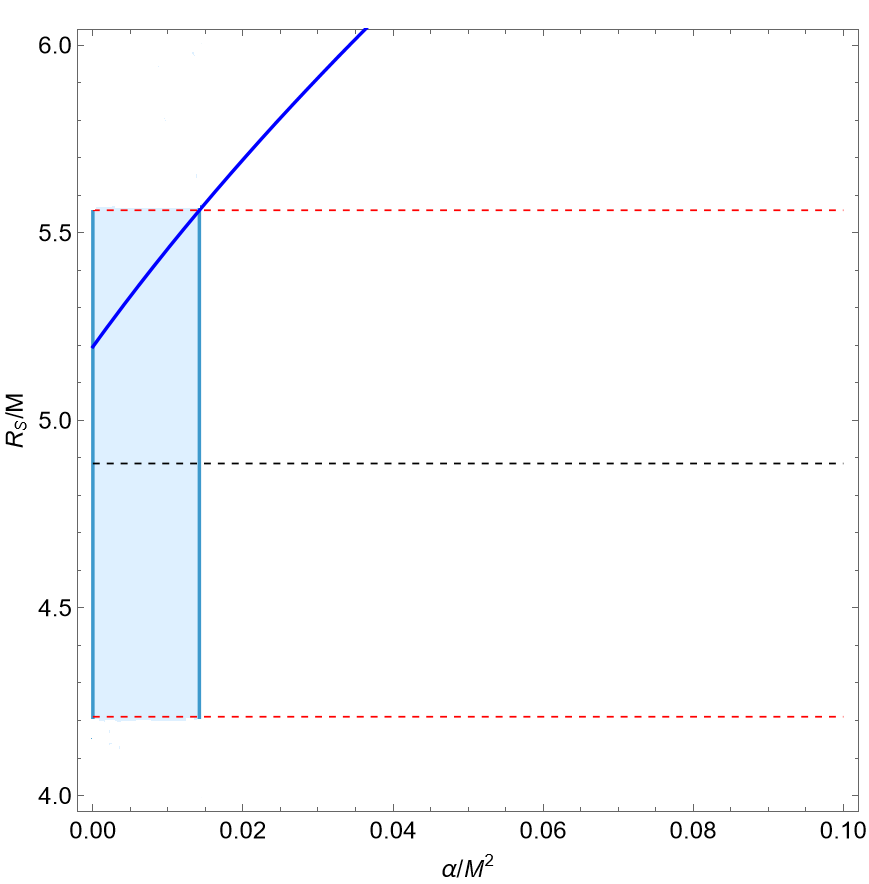}\\
  \includegraphics[width=0.75 \columnwidth]{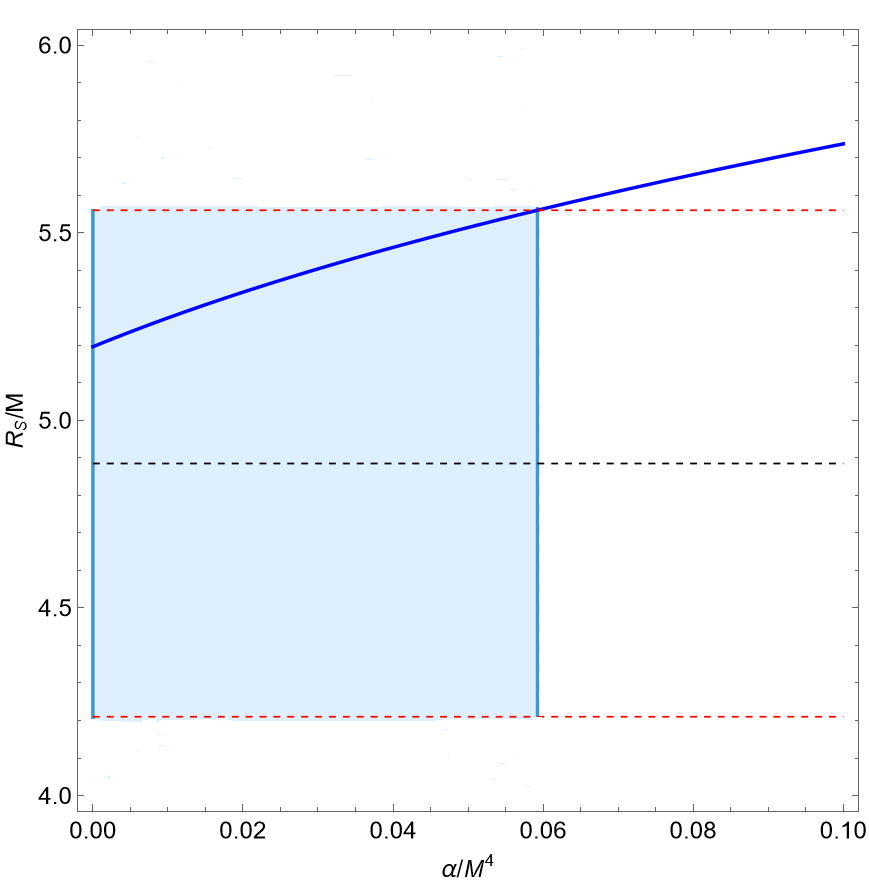}
  \caption{{\it{The allowed range of $\alpha/M^2$ (upper panel), and 
$\alpha/M^4$ 
(lower panel) for Case II black-hole solution with $n=2, 3$ in light of EHT 
diameter measurement of  Sgr A*. The black-dashed line represents the central 
value of $R_s/M$ in EHT data, while the red-dashed lines mark the lower and 
upper permissible bounds. } }}
  \label{f4}
 \end{figure}

 The case of Sgr~A* is more challenging due to variability and scattering 
 effects, but using the mass-to-distance ratio inferred from Keck and VLTI 
 measurements, Ref.~\cite{Vagnozzi:2022moj} reports a $2\sigma$ interval:
 \begin{equation}
  4.21 \lesssim \frac{R_s}{M} \lesssim 5.56.
 \end{equation}
 This bound can similarly be used to constrain the shift in the shadow 
radius 
 induced by~$\alpha$. Substituting the theoretical expressions for $R_s^{\rm 
  (II)}$ into this interval yields an independent and complementary 
constraint 
 on~$\alpha$,  illustrated in Fig.~\ref{f4}. Although the Sgr~A* bounds are 
generally weaker than those from 
 M87*, they remain consistent with the same allowed intervals and exclude 
 significantly larger deviations from GR.
 
 \subsection{Summary of EHT constraints}

 Taken together, the M87* and Sgr~A* observations place robust constraints 
on  the deformation parameter of the Case~II covariant $f(Q)$ black hole. The 
bounds  obtained from M87*, owing to its larger mass and better-defined image 
structure,  are the most stringent, bounding the fractional corrections of 
order  $\alpha/M^2$ to below a few percent in the $n=2$ case, and placing 
meaningful  constraints on $\alpha/M^4$ for $n=3$. These results demonstrate 
that the  current EHT data already probe nonmetricity-induced deviations from 
GR 
at the  horizon scale, and thus future higher-resolution observations will 
further tighten 
 these limits.

\section{Constraints from S2 star precession}
 \label{s2}
 
 The stellar dynamics of the Galactic Center provide one of the cleanest 
tests of 
 gravity in the weak-to-intermediate field regime surrounding a supermassive 
 black hole. Among the stars orbiting Sgr~A*, the S2 star (also known as 
S0-2) 
 plays a central role. It is a bright, massive B-type star with a short 
orbital 
 period ($\sim 16$~yr) and an exceptionally small pericenter distance ($\sim 
 120$~AU). Its highly eccentric orbit, combined with precise astrometric and 
 spectroscopic measurements from  GRAVITY Keck+VLTI
\cite{GRAVITY:2021xju}, makes S2 an 
outstanding 
 natural probe of deviations from the Schwarzschild geometry of Sgr~A*.
 
 In Newtonian gravity, the orbit of a test mass is a closed ellipse, while in 
GR, the curvature of spacetime induces a forward rotation of 
 the orbit, leading to a shift in the periapsis after each revolution. This 
 periastron precession is one of the classical relativistic tests and is 
directly 
 sensitive to the underlying spacetime metric. Hence, any modification to GR, 
such as the 
 $\alpha$-dependent corrections predicted by covariant $f(Q)$ gravity, 
modifies the precession rate.
 
 In this section we use the observational measurements of S2’s periastron 
 precession to derive upper bounds on the deformation parameter $\alpha$ for 
both 
 Case~I and Case~II black hole geometries. The modified precession 
expressions 
 based on the solutions of Ref.~\cite{Wang:2024dkn} are
 \begin{eqnarray}
&&
\!\!\!\!\!\!\!\!\!\!\!\!\!\!\!\!\!\!\!
\Delta\phi_{\rm n=2}^{I} \approx \frac{6\pi M}{a(1-e^{2})} + 
\frac{\pi\Lambda 
   a^{3}(1-e^{2})^{3}}{M}
   \nonumber\\
&& \ \ \ \ 
  + \alpha\!\left[\frac{16\pi M^{2}}{a^{4}(1-e^{2})^{4}} + \frac{40\pi 
   M^{3}}{a^{5}(1-e^{2})^{5}}\right], \label{In2}  \\
&&\!\!\!\!\!\!\!\!\!\!\!\!\!\!\!\!\!\!\!
\Delta\phi_{\rm n=3}^{I} \approx \frac{6\pi M}{a(1-e^{2})} + 
\frac{\pi\Lambda 
   a^{3}(1-e^{2})^{3}}{M} \nonumber\\
&& \ \ \ \ 
  - \alpha\!\left[\frac{96\pi M^{4}}{a^{8}(1-e^{2})^{8}} + \frac{168\pi 
   M^{5}}{a^{9}(1-e^{2})^{9}}\right], \label{In3}
  \\
&&\!\!\!\!\!\!\!\!\!\!\!\!\!\!\!\!\!\!\! \Delta\phi_{\rm n=2}^{II} \approx 
\frac{6\pi M}{a(1-e^{2})} + \frac{\pi\Lambda 
   a^{3}(1-e^{2})^{3}}{M} \nonumber\\
&& \ \ \ \ 
  + \alpha\!\left[\frac{32\pi}{a(1-e^{2})M} + 
  \frac{256\pi}{a^{2}(1-e^{2})^{2}}\right], \label{IIn2}
  \\
&&\!\!\!\!\!\!\!\!\!\!\!\!\!\!\!\!\!\!\!  \Delta\phi_{\rm n=3}^{II} \approx 
\frac{6\pi M}{a(1-e^{2})} + \frac{\pi\Lambda 
   a^{3}(1-e^{2})^{3}}{M} \nonumber\\
&& \ \ \ \ 
  + \alpha\!\left[\frac{512\pi}{a^{3}(1-e^{2})^{3}M} + 
  \frac{2048\pi}{a^{4}(1-e^{2})^{4}}\right].
  \label{IIn3}
 \end{eqnarray}

 The first term in each line is the standard Schwarzschild prediction, while 
terms 
 proportional to $\alpha$ encode deviations arising from the covariant 
$f(Q)$ 
 model. On Galactic scales, the cosmological constant term is negligible and as 
mentioned above it is omitted in our analysis.
 
 To constrain $\alpha$, we compare the theoretical prediction
 \begin{equation}
  \Delta\phi_{f(Q)} = \Delta\phi_{\rm GR} + \delta\phi_{f(Q)}(\alpha),
 \end{equation}
 with the observational result
 \begin{equation}
  \Delta\phi_{\rm obs} = \Delta\phi_{\rm GR} + \delta\phi_{\rm obs},
 \end{equation}
 where $\delta\phi_{\rm obs}$ quantifies observational uncertainties or any 
 potential deviation from GR. Requiring consistency between theory and 
 observation imposes
 \begin{equation}
  \label{obs}
  \big|\delta\phi_{f(Q)}(\alpha)\big| \lesssim \big|\delta\phi_{\rm 
obs}\big|.
 \end{equation}
  
 We adopt the following observationally inferred parameters:
 \begin{itemize}
  \item Mass of Sgr~A*:  
  \(M = 4.261 \times 10^{6} M_{\odot}\),  
  corresponding to \(M \approx 6.29 \times 10^{6}~{\rm km}\) in geometric 
units.
  \item Semi-major axis:  
  using Keck + VLTI distance \(R_0 = 8.275~{\rm kpc}\),  
  the observed angular size \(a = 0.125''\) gives  
  \(a \approx 1.54 \times 10^{14}~{\rm km}\).
  \item Orbital eccentricity: \(e \approx 0.884\).
  \item Observed precession uncertainty:  
  GRAVITY finds consistency with GR at the $\sim$10\% level, yielding  
  \(\delta\phi_{\rm obs} \approx 3.52 \times 10^{-7}~{\rm rad/orbit}\).
 \end{itemize}
 Moreover, we define
 \begin{equation}
  u = \frac{1}{a(1-e^{2})},
 \end{equation}
 which for S2 evaluates to  
 \(u \approx 2.97 \times 10^{-14}~{\rm km^{-1}}\).  
 This quantity conveniently packages the strong dependence of the precession 
on 
 the orbital size and eccentricity.

 We proceed by  inserting S2’s orbital parameters into the $\alpha$-dependent 
contributions 
 to the precession,  and we impose the bound~\eqref{obs}.\\
 
 \paragraph*{Case I with $n=2$.}
 From Eq.~\eqref{In2} we find
 \[
 \delta\phi_{f(Q)}^{I,n=2} \approx \alpha\!\left[16\pi M^{2} u^{4} + 40\pi 
M^{3} 
 u^{5}\right].
 \]
 The second term is suppressed by seven orders of magnitude and can be 
neglected. 
 Numerically,
 \[
 |\delta\phi_{f(Q)}^{I,n=2}| \approx |\alpha| \cdot 1.55\times10^{-40},
 \]
 leading to
 \[
 |\alpha| \lesssim 2.27 \times 10^{33}~{\rm km}^{2}.
 \]
 \\
 \paragraph*{Case I with $n=3$.}
 From Eq.~\eqref{In3} we have
 \[
 \delta\phi_{f(Q)}^{I,n=3} \approx -\alpha\!\left[96\pi M^{4} u^{8}\right],
 \]
 with the subleading term again negligible. We obtain
 \[
 |\delta\phi_{f(Q)}^{I,n=3}| \approx |\alpha| \cdot 2.11\times10^{-80},
 \]
 and therefore
 \[
 |\alpha| \lesssim 1.67 \times 10^{73}~{\rm km}^{4}.
 \]
 \\
 \paragraph*{Case II with $n=2$.}
 From Eq.~\eqref{IIn2} we have
 \[
 \delta\phi_{f(Q)}^{II,n=2} \approx \alpha\!\left[32\pi \frac{u}{M} \right],
 \]
 where the second term is negligible. Numerically we find
 \[
 |\delta\phi_{f(Q)}^{II,n=2}| \approx |\alpha| \cdot 4.74\times10^{-19},
 \]
 yielding a much tighter bound:
 \[
 |\alpha| \lesssim 7.42 \times 10^{11}~{\rm km}^{2}.
 \]
 
 \paragraph*{Case II with $n=3$.}
 From Eq.~\eqref{IIn3} we find
 \[
 \delta\phi_{f(Q)}^{II,n=3} \approx \alpha\!\left[512\pi 
\frac{u^{3}}{M}\right],
 \]
 leading to
 \[
 |\delta\phi_{f(Q)}^{II,n=3}| \approx |\alpha| \cdot 6.71\times10^{-45},
 \]
 and therefore
 \[
 |\alpha| \lesssim 5.25 \times 10^{37}~{\rm km}^{4}.
 \]

  \begin{table}[ht]
  \centering
  \caption{Constraints on $\alpha$-parameter from S2 star precession data.}
  \label{tab:comprehensive_bounds}
  \begin{tabular}{cccc}
   \hline
   Case  &   $\ \ \ n\ \ \ $  & $\  $ Dimensional Bound  $\  $ & 
  Dimensionless Bound  \\
   \hline
   I & 2 & $|\alpha| \lesssim 10^{33}$ km$^2$ & 
  $\dfrac{|\alpha|}{M^2}\lesssim 10^{20}$ \vspace{0.2cm} \\ \vspace{0.2cm}
    I  & 3 & $|\alpha| \lesssim 10^{73}$ km$^4$ &   
$\dfrac{|\alpha|}{M^4}\lesssim 10^{46}$ \\ \vspace{0.2cm}
   II  & 2 & $|\alpha| \lesssim 10^{12}$ km$^2$ & 
  $\dfrac{|\alpha|}{M^2}\lesssim 10^{-2}$ \\ \vspace{0.1cm}
     II  & 3 & $|\alpha| \lesssim 10^{38}$ km$^4$ &  
$\dfrac{|\alpha|}{M^4}\lesssim 10^{10}$ \\
   \hline
  \end{tabular}
 \end{table}

 In summary, as we see, the S2 constraints are naturally much weaker than 
those 
 derived from the EHT shadow measurements, since the S2 orbit probes the 
 spacetime at scales $r \sim 10^{13}$--$10^{14}$~km, far beyond the photon 
 sphere. Nevertheless, they provide a valuable complementary test in a 
different 
 dynamical regime and are fully consistent with the bounds obtained from M87* 
and 
 Sgr~A* imaging. Expressing the constraints in dimensionless combinations 
such as 
 $\alpha/M^{2}$ or $\alpha/M^{4}$ enhances their interpretability. 
 Table~\ref{tab:comprehensive_bounds} summarizes the complete set of bounds.

 \section{Strong lensing constraints}
 \label{len}
 
 Gravitational lensing in the strong-field regime provides a powerful and 
highly  sensitive probe of the spacetime geometry near compact objects, making 
it an  excellent observational test for modified gravity theories. In the 
context of  covariant $f(Q)$ gravity, deviations from GR modify 
both the  location of the photon sphere and the behavior of light deflection 
near it,  leading to potentially observable signatures. In this section we 
review the  strong-deflection formalism and we apply it to the black-hole 
solutions obtained above, in order to quantify their impact on lensing 
observables and derive constraints on the parameter $\alpha$.

 \subsection{Strong-field deflection angle and Bozza formalism}

 The theoretical framework for strong-field gravitational lensing was 
established
 through a series of key developments. The pioneering work of Virbhadra and 
Ellis
 (VE)~\cite{Virbhadra:2002ju} shifted the focus from weak-field deflections 
to
 the strong-field regime by numerically showing that light rays performing
 multiple loops around a compact object generate an infinite sequence of
 relativistic images. Subsequently, Frittelli, Kling, and Newman
 (FKN)~\cite{Frittelli:1999yf} placed the subject on a rigorous mathematical
 footing by formulating lensing observables directly from the geometry of the
 observer’s past light cone, without invoking weak-field approximations. 
Building
 on these advances, Bozza~\cite{Bozza:2002zj} derived a universal analytic
 expression for the deflection angle in the strong-field limit. By carefully
 expanding the deflection integral near the photon sphere, he demonstrated 
that
 its divergence is logarithmic and universal, thereby providing a practical 
and
 powerful method to compute lensing observables for any spherically symmetric
 spacetime that possesses a photon sphere.

 In particular, in Ref.~\cite{Virbhadra:2002ju}, it was shown that for a 
static, spherically
 symmetric metric of the form
 \begin{equation}
  \label{gen}
  ds^2 = -A(r)\,dt^2 + B(r)\,dr^2 + C(r)\,(d\theta^2 + 
\sin^2\theta\,d\phi^2),
 \end{equation}
 the deflection angle $\hat{\beta}(r_0)$ of a light ray with closest approach
 radius $r_0$ is
 \begin{equation}
  \hat{\beta}(r_0)
  =
  -\pi
  + 2 \int_{r_0}^{\infty}
  \frac{dr}{
   \sqrt{ \dfrac{C(r)}{B(r)} }
   \sqrt{ \dfrac{A(r_0)C(r)}{A(r)C(r_0)} - 1 } } \,.
 \end{equation}
 In the strong deflection limit $r_0 \to r_c^{+}$, where $r_c$ is the photon
 sphere radius, this integral diverges logarithmically.
 
 A more explicit and widely used derivation of the logarithmic expansion was
 obtained by Bozza~\cite{Bozza:2002zj}, whose formalism has become the 
standard
 reference for strong-field lensing calculations. For the metric
 \eqref{gen}, the deflection angle for impact parameters $b$ close to the
 critical value $b_c$ can be written as
 \begin{equation}
  \label{def}
  \beta_{\text{def}}(b)
  =
  -\bar{a} \ln\!\left( \frac{b}{b_c} - 1 \right)
  + \bar{b}
  + \mathcal{O}(b - b_c),
 \end{equation}
 where $\bar{a}$ and $\bar{b}$ are the strong-deflection coefficients, and 
$r_c$
 and $b_c$ are the photon sphere radius and the corresponding critical impact
 parameter, respectively.
 
 In order to compute $\bar{a}$ and $\bar{b}$, one introduces the function
 \begin{equation}
  R(r, r_0)
  =
  \sqrt{ \frac{C(r)\, B(r)}{C(r_0)\, B(r_0)} }
  \left(
  \frac{A(r_0)}{A(r)} - \frac{C(r_0)}{C(r)}
  \right),
 \end{equation}
and therefore \cite{Bozza:2002zj}
 \begin{eqnarray}
  &&\!\!\!\!\!\!\!\!\!\!\!\!\!\!\!
  \bar{a}
  =
  \sqrt{
   \frac{2 A(r_c) C(r_c)}
   {C''(r_c) A(r_c) - A''(r_c) C(r_c)}
  },
  \label{a}
  \\
 &&\!\!\!\!\!\!\!\!\!\!\!\!\!\!\! \bar{b}
  =
  -\pi + b_R\nonumber\\
 &&\!\!\!\!\!\!\!\!\!\!\!\!\! \!\! + \bar{a}\,
  \ln\!\left(\!
  \frac{2 C(r_c)^2 A'(r_c)^2}
  {A(r_c)^3 C(r_c)\,[C(r_c) A''(r_c) \!-\! A(r_c) C''(r_c)]}
 \! \right),
  \label{bb}
 \end{eqnarray}
 where the regular part $b_R$ is given by
 \begin{equation}
  b_R
  =
  2 \bar{a} \int_0^1
  \left[
  \frac{1}{\sqrt{ R(z, r_c)\, f(z) }}
  -
  \frac{1}{\bar{a} z}
  \right] dz,
 \end{equation}
 with the change of variable $z = 1 - r_c/r$ and
 \begin{equation}
  f(z)
  =
  \frac{1 - A(r_c)/A(r)}
  {C(r_c)/C(r) - A(r_c)/A(r)}.
 \end{equation}
 Two remarks are worth noting here. Firstly, the coefficient $\bar{a}$ controls 
the
 strength of the logarithmic divergence as $b \to b_c^+$, and for the 
Schwarzschild
 black hole, one finds $\bar{a} = 1$. Secondly, the coefficient $\bar{b}$ 
encodes
 the finite offset of the deflection angle, and for Schwarzschild solution we 
have
 $\bar{b} = \ln[216(7-4\sqrt{3})] - \pi \approx -0.4002$.
 
 Using  \eqref{a}, \eqref{bb}, and \eqref{def}, for the Case~I black hole
 solutions discussed in Sec.~\ref{BHsolutions} we obtain the following
 $\alpha$-corrected expressions:\\ 
 \paragraph*{Case I with $n=2$.}
\begin{eqnarray}
&   &
\!\!\!\!\!\!\!\!\!\!\!\! \bar{a} \approx 1 - 0.07243 \frac{\alpha}{M^2}, 
\nonumber\\
 &   &
\!\!\!\!\!\!\!\!\!\!\!\!  \bar{b} \approx -0.40023 + 0.152 \frac{\alpha}{M^2}, 
\nonumber \\
  &   &
\!\!\!\!\!\!\!\!\!\!\!\!  \hat{\beta}_{\text{def}}(b)
  \approx
   -\left(1 - 0.07243 \frac{\alpha}{M^2}\right)
   \ln\!\left( \frac{b}{b_c} - 1 \right)\nonumber\\
 & & \ \ \ \ \ \  \, 
   - 0.40023 + 0.152 \frac{\alpha}{M^2}.
        \label{In2}
  \end{eqnarray}
 \\ 
 \paragraph*{Case I with $n=3$.}
\begin{eqnarray}
&   &
\!\!\!\!\!\!\!\!\!\!\!\! 
\bar{a} \approx 1 + 0.00176 \frac{\alpha}{M^4},\nonumber \\
 &   &
\!\!\!\!\!\!\!\!\!\!\!\!  \bar{b} \approx -0.40023 - 0.00342 \frac{\alpha}{M^4}, 
\nonumber \\
  &   &
\!\!\!\!\!\!\!\!\!\!\!\!  
\hat{\beta}_{\text{def}}(b)
   \approx
   -\left(1 + 0.00176 \frac{\alpha}{M^4}\right)
   \ln\!\left( \frac{b}{b_c} - 1 \right)\nonumber\\
 & & \ \ \ \ \ \  \, 
   - 0.40023 - 0.00342 \frac{\alpha}{M^4}.
        \label{In3}
  \end{eqnarray}
 
 As we can see, for Case~I with $n=2$, the $\alpha$-correction appears at 
order $1/M^2$, 
making
 it in principle more relevant for stellar-mass black holes than for
 supermassive ones. A positive $\alpha$ slightly reduces $\bar{a}$ (weaker
 divergence) and increases $\bar{b}$ (larger offset), leading to small but
 well-defined shifts in the positions and magnifications of relativistic 
images.
 For $n=3$, the corrections are suppressed by $1/M^4$ and become effectively
 negligible for astrophysical black holes, rendering this case extremely 
close to
 the Schwarzschild limit.
 
 Performing the same analysis for the Case~II black hole solutions yields:
\\
 \paragraph*{Case II with $n=2$.}
 \begin{eqnarray}
&   &
\!\!\!\!\!\!\!\!\!\!\!\! 
\bar{a} \approx 1 - 9.244 \frac{\alpha}{M^2}, \nonumber\\
 & & \!\!\!\!\!\!\!\!\!\!\!\!  \bar{b} \approx -0.40023 + 18.5 
\frac{\alpha}{M^2},\nonumber \\
 && \!\!\!\!\!\!\!\!\!\!\!\!  \hat{\beta}_{\text{def}}(b)
   \approx
   -\left(1 - 9.244 \frac{\alpha}{M^2}\right)
   \ln\!\left( \frac{b}{b_c} - 1 \right)\nonumber\\
  & & \ \ \ \ \ \  \, 
   - 0.40023 + 18.5 \frac{\alpha}{M^2}.
     \label{IIn2}
  \end{eqnarray}
 \\
 \paragraph*{Case II with $n=3$.}
 \begin{eqnarray}
&   &
\!\!\!\!\!\!\!\!\!\!\!\! 
   \bar{a} \approx 1 - 0.842 \frac{\alpha}{M^4},\nonumber \\
&   &
\!\!\!\!\!\!\!\!\!\!\!\! \bar{b} \approx -0.40023 + 1.523 
\frac{\alpha}{M^4},\nonumber \\
 &   &
\!\!\!\!\!\!\!\!\!\!\!\!  \hat{\beta}_{\text{def}}(b)
   \approx
   -\left(1 - 0.842 \frac{\alpha}{M^4}\right)
   \ln\!\left( \frac{b}{b_c} - 1 \right)\nonumber\\
  & & \ \ \ \ \ \  \, 
   - 0.40023 + 1.523 \frac{\alpha}{M^4}.
  \label{IIn3}
 \end{eqnarray}
 
 As we see, for the same value of $n$, Case~II exhibits much larger 
corrections than
 Case~I, reflecting the different radial dependence of the corresponding 
metric
 functions. In particular, the $n=2$ Case~II solution shows pronounced
 deviations: a positive $\alpha$ substantially decreases $\bar{a}$ (weaker
 divergence) and strongly enhances $\bar{b}$ (larger offset), with 
corrections
 roughly two orders of magnitude larger than in Case~I, $n=2$. The $n=3$ 
Case~II  corrections are smaller than in the $n=2$ case, but they remain 
significantly larger
 than their Case~I counterparts.
 
 In summary, these strong-lensing results indicate that Case~II solutions in
 covariant $f(Q)$ gravity, especially for $n=2$, can lead to observationally
 relevant modifications in the strong deflection regime. Through their impact 
on  the strong-deflection coefficients $\bar{a}$ and $\bar{b}$, they alter the
 positions, separations, and magnifications of relativistic images, making 
these  models promising targets for future high-resolution observations of 
black hole  shadows and strong-field lensing with next-generation instruments 
building upon  the Event Horizon Telescope.

 \subsection{Shedding light on relativistic images with EHT data}
 
 In the strong-deflection regime, the deflection angle is fundamentally a 
 function of the impact parameter \(b\), as given in (\ref{def}). For a 
 distant observer located at a distance \(D_{OL}\) from the black hole, the 
 impact parameter is related to the observed angular position \(\theta\) 
through 
 the standard small-angle relation
 \be \label{D}
 b = \theta\, D_{OL},
 \ee
 which follows from geometric optics and the fact that the angular size of a 
 supermassive black hole is of the order of tens of microarcseconds.  
 Substituting  (\ref{D}) into the strong-deflection formula (\ref{def}) 
gives
 \be\label{3}
 \beta(\theta) = -\bar{a}\,
 \ln\!\left(\frac{\theta D_{OL}}{b_c}-1\right) 
 + \bar{b} + \mathcal{O}(\theta D_{OL}-b_c).
 \ee
 
 The lens equation for an aligned source (\(\psi = 0\)) is
 \be
 \beta(\theta) = 2\pi n,
 \ee
 where \(n\) counts the number of loops performed by the photon around the 
black 
 hole.  
 Solving Eq.~(\ref{3}) for the angular position of the \(n\)-th relativistic 
 image yields
 \be
 \theta_n D_{OL} 
 = 
 b_c\left[1 + e^{\frac{\bar{b}-2\pi n}{\bar{a}}}\right].
 \ee
 Then, introducing the asymptotic image position \(\theta_\infty=b_c/D_{OL}\), 
one 
 recovers the Bozza’s formula, namely
 \be \label{5}
 \theta_n = \theta_\infty\left[1 + e^{\frac{\bar{b}-2\pi n}{\bar{a}}}\right],
 \ee
 showing that all relativistic images cluster exponentially close to 
 \(\theta_\infty\).  
 
 Of particular interest is the first relativistic image (\(n=1\)), whose 
position 
 is
 \be\label{55}
 \theta_1 = \theta_\infty\left[1 + e^{\frac{\bar{b}-2\pi}{\bar{a}}}\right].
 \ee
 Since in GR, the asymptotic image location is
 \be\label{GR}
 \theta_\infty^{\rm GR} 
 = \frac{6\sqrt{3}\,GM}{c^2 D_{OL}},
 \ee
 which corresponds to the angular radius of the Schwarzschild photon sphere, 
any deviation in the strong-field coefficients \(\bar{a}\) and \(\bar{b}\) 
 induced by \(f(Q)\) gravity modifies both \(\theta_\infty\) and the 
exponential  clustering of relativistic images.

 Finally, let us come back to the  EHT data. As we saw above, for M87* 
  and Sgr~A*, the measured angular diameters are respectively
 \[
 \theta_{d,\mathrm{M87*}} = (42 \pm 3)\,\mu\mathrm{as}, 
 \qquad
 \theta_{d,\mathrm{SgrA*}} = (48.7 \pm 7)\,\mu\mathrm{as}.
 \]
 The GR predictions obtained from Eq.~(\ref{GR}) are
 \[
 \theta_{\infty,\mathrm{M87*}}^{0} \simeq 39.68\,\mu\mathrm{as}, 
 \qquad
 \theta_{\infty,\mathrm{SgrA*}}^{0} \simeq 52.8\,\mu\mathrm{as}.
 \]
 Since EHT observations are fully consistent with GR within uncertainties, 
any 
 modification to the asymptotic image location must satisfy
 \be
 \left|\frac{\Delta \theta_\infty}{\theta_\infty^{0}}\right|
 <\frac{\text{EHT uncertainty}}{\theta_{d,\mathrm{GR}}}.
 \ee
 At the 
 \(2\sigma\) level this implies the bounds
 \be\label{bound}
 \left|\frac{\Delta \theta_\infty}{\theta_\infty^{0}}\right|_{\mathrm{M87*}}
 \lesssim 0.15,
 \qquad
 \left|\frac{\Delta \theta_\infty}{\theta_\infty^{0}}\right|_{\mathrm{SgrA*}}
 \lesssim 0.26.
 \ee
  Using the strong-deflection coefficients for Case I and Case II,
 namely  (\ref{In2})-(\ref{IIn3}), and inserting them into 
expression (\ref{55}), we 
obtain 
 the constraints on the dimensionless parameters
$
 \frac{\alpha}{M^{2n-2}}$
 summarized in Table~\ref{tab:alpha_bounds}.  
 We find that Case~II is constrained significantly more tightly, by roughly 
three 
 orders of magnitude, comparing to Case~I.  
 This reflects the fact that Case~II induces much larger corrections to the 
 photon-sphere radius and strong-deflection coefficients, making it far more 
 sensitive to current EHT observations and a promising target for 
next-generation 
 horizon-scale interferometry.
 
   \begin{table}[h!]
  \centering
  \caption{Constraints on $\alpha$-parameter from strong-field deflection and  
EHT
data.}
  \label{tab:alpha_bounds}
  \begin{tabular}{cccc}
  \hline
   Case & $\ \ \ n\ \ \ $ &$\ \ $ Source $\ \ $ &$\ \ $ Upper Bound of 
$\left|\dfrac{\alpha}{M^{2n-2}}\right|$ \\
  \hline   
   I  & 2 & M87* & $\lesssim  
30.36$ \\
   I  & 2 & Sgr A* & $\lesssim 52.63$\vspace{0.1cm} \\
       I  & 3 & M87* & $\lesssim 
738.92$ \\
      I  & 3 & Sgr A* & $\lesssim 1280.79$\vspace{0.1cm} \\
   \midrule
   II & 2 & M87* & $\lesssim 
0.02812$ \\
   II & 2  & Sgr A* & $\lesssim 0.04875$\vspace{0.1cm} \\
  II & 3 & M87* & $\lesssim 
0.09494$ \\
  I  & 3 & Sgr A* & $\lesssim 0.1646$ \\
        \hline
  \end{tabular}
   \end{table}

 \section{Conclusions}\label{con}

 Modified gravity continues to play a central role in addressing open questions 
 in gravitational physics and cosmology, ranging from the nature of dark 
energy  to the ultraviolet completion of General Relativity (GR). Among the 
various geometric reformulations of gravity, covariant symmetric teleparallel 
gravity  offers a compelling framework in which 
non-metricity,  rather than curvature or torsion, carries the gravitational 
interaction. The  promotion of the non-metricity scalar $Q$ to a general 
function $f(Q)$ enriches  the structure of the theory, allowing new 
phenomenology at both cosmological  and  astrophysical scales, while preserving 
second-order field equations. These 
 features make $f(Q)$ gravity an attractive and well-motivated arena for 
 exploring  deviations from GR.
 
 Testing modified gravity requires pushing beyond the weak-field and 
 cosmological  limits into the strong-gravity regime, where nonlinear effects 
become dominant 
 and deviations, if present, can be amplified. Black holes, in particular, 
serve  as 
 exceptional laboratories due to their simplicity, universality, and the 
extreme curvature near their horizons. The advent of horizon-scale observations 
such as 
 the EHT images of M87* and Sgr~A*, measurements of the S2 star orbit around 
the  Galactic Center, and the development of precise strong-lensing frameworks 
have opened new pathways for probing gravity in conditions inaccessible by any 
other  means. By combining shadow sizes, photon-sphere properties, periastron 
 precession, and strong-field lensing coefficients, it becomes possible to 
place  simultaneous, independent, and complementary constraints on viable 
extensions of GR.
 
 In this work, we performed the first unified strong-field analysis of 
 power-law $f(Q)$ gravity using spherically symmetric black-hole solutions, 
 generated by distinct classes of symmetric affine connections. For each 
 solution, we studied the behavior of null geodesics, we derived analytic 
 expressions for the photon-sphere radius and shadow size, and we calculated 
 strong-lensing observables within the Bozza formalism. We showed that the 
 two families of solutions, labeled here as Case~I and Case~II, exhibit 
markedly  different sensitivity to the deformation parameter $\alpha$, with 
Case~II  displaying significantly stronger departures from GR for the same 
magnitude of 
 $\alpha$.
 
 We further confronted these predictions with a broad suite of strong-field 
 observables. The EHT shadow diameters of M87* and Sgr~A* yield tight bounds 
on  the allowed deviations in the photon-sphere radius; the S2 star precession 
 provides independent and complementary constraints on the same parameter 
space; 
 and the strong-lensing coefficients $\bar{a}$ and $\bar{b}$ impose 
additional 
 restrictions through the logarithmic structure of the deflection angle. 
When 
 combined, these probes consistently indicate that Case~I solutions remain 
very 
 close to GR for astrophysical black holes, whereas Case~II solutions are 
 significantly constrained, especially for the $n=2$ power-law model where 
 deviations become observationally significant.
 
 Looking ahead, the rapidly improving landscape of strong-field observations 
 promises even more decisive tests of $f(Q)$ gravity. Future EHT campaigns 
with  enhanced arrays, high-frequency imaging, and time-domain capabilities 
will  sharply reduce uncertainties in shadow shapes. Next-generation 
astrometric  missions will measure stellar orbits around Sgr~A* with 
unprecedented  precision,  and upcoming facilities such as the ngEHT and 
space-based interferometers will 
 open access to relativistic images and higher-order lensing effects. On the 
 theoretical side, extending the present analysis to rotating black-hole 
 solutions, dynamical spacetimes, and multimessenger signatures will allow 
 broader classes of $f(Q)$ models to be tested. Together, these developments 
 have  the potential to transform strong-field astrophysics into a definitive 
tool  for probing non-metricity-based modifications of gravity.

 \begin{acknowledgments}
  ENS acknowledges the contribution of the LISA   CosWG, and of   COST   Actions 
 CA18108  ``Quantum Gravity Phenomenology in the multi-messenger   approach''  
and  CA21136 ``Addressing observational tensions in cosmology with 
  systematics and fundamental physics (CosmoVerse)''. 
 \end{acknowledgments}
 
 
\end{document}